\documentclass[aps,twocolumn,amsmath,amssymb,prb,showpacs,superscriptaddress]{revtex4-1}
\usepackage{amsfonts, hyperref, times}
\usepackage{graphicx}

\providecommand{\vect}[1]{{\boldsymbol{#1}}}

\begin{document}

\title{Characterizing breathing dynamics of magnetic (anti-)skyrmions within the Hamiltonian formalism}

\author{B.F. McKeever}
\email{bmckeever@uni-mainz.de}
\affiliation{Institute of Physics, Johannes Gutenberg-Universit{\"a}t, 55128 Mainz, Germany}
\affiliation{Graduate School Materials Science in Mainz, Staudingerweg 9, 55128 Mainz, Germany}
\author{D.R. Rodrigues}
\affiliation{Institute of Physics, Johannes Gutenberg-Universit{\"a}t, 55128 Mainz, Germany}
\affiliation{Graduate School Materials Science in Mainz, Staudingerweg 9, 55128 Mainz, Germany}
\author{D. Pinna}
\affiliation{Institute of Physics, Johannes Gutenberg-Universit{\"a}t, 55128 Mainz, Germany}
\author{Ar.~Abanov}
\affiliation{Department of Physics \& Astronomy, Texas A\&M University, College Station, Texas 77843-4242, USA}
\author{Jairo Sinova}
\affiliation{Institute of Physics, Johannes Gutenberg-Universit{\"a}t, 55128 Mainz, Germany}
\affiliation{Institute of Physics ASCR, v.v.i, Cukrovarnicka 10, 162 00 Prag 6, Czech Republic}
\author{K. Everschor-Sitte}
\affiliation{Institute of Physics, Johannes Gutenberg-Universit{\"a}t, 55128 Mainz, Germany}
\date{\today}

\begin{abstract}
We derive an effective Hamiltonian system describing the low energy dynamics of circular magnetic skyrmions and antiskyrmions. 
Using scaling and symmetry arguments we model (anti-)skyrmion dynamics through a finite set of coupled, canonically conjugated, collective coordinates.
The resulting theoretical description is independent of both micromagnetic details as well as any specificity in the ansatz of the skyrmion profile. 
Based on the Hamiltonian structure we derive a general description for breathing dynamics of (anti-)skyrmions in the limit of radius much larger than the domain wall width. 
The effective energy landscape reveals two qualitatively different types of breathing behavior. 
For small energy perturbations we reproduce the well-known small breathing mode excitations, where the magnetic moments of the skyrmion oscillate around their equilibrium solution. 
At higher energies we find a breathing behavior where the in-plane angle of the skyrmion continuously precesses, transforming N\'eel to Bloch skyrmions and vice versa. 
For a damped system we observe the transition from the continuously rotating and breathing skyrmion into the oscillatory one. 
We analyze the characteristic frequencies of both breathing types, as well as their amplitudes and distinct energy dissipation rates. 
For rotational (oscillatory) breathing modes we predict on average a linear (exponential) decay in energy. 
We argue that this stark difference in dissipative behavior should be observable in the frequency spectrum of excited (anti-)skyrmions.
\end{abstract}


\maketitle

\section{Introduction}
\label{sec:Intro}
Topological magnetic textures have attracted substantial attention in spintronics~\cite{Wolf2006,Chappert2007,Jonietz2010, Iwasaki2013a} in light of prospects to harness their favorable properties for magnetic memory technologies,~\cite{Parkin2008,Nagaosa2013, Fert2013} information processing~\cite{Zhang2015l, Pinna2017, Luo2018, Zazvorka, Chauwin2018} and novel approaches to computation.~\cite{Huang2016b, He2017a, Li2017a, He2017, Prychynenko2017, Azam2018, Bourianoff2018}
An important building block in this direction is understanding their dynamical excitations in order to test stability and to devise ways to efficiently manipulate them. 
Among the magnetic textures studied are domain walls~\cite{Yamanouchi2004,Tatara2004,Hayashi2006,Parkin2008} and skyrmions.~\cite{Bogdanov1989,Bogdanov1994,Muhlbauer2009a, Schulz2012, Iwasaki2013a, Sampaio2013, Nagaosa2013}
With respect to possible applications in memory devices, skyrmions present several advantages over domain walls due to their smaller sizes, lower threshold for current driven mobility,~\cite{Jonietz2010,Yu2012,Schulz2012,Fert2013} and their tendency to avoid obstacles and boundaries.~\cite{Zhang2015c}
Recently, a growing collection of other exotic relatives are also receiving attention.
This includes magnetic solitons of higher topological order,~\cite{Zhang2016i} chiral bobbers,~\cite{Zheng2017} non-topological counterparts such as the skyrmionium particle,~\cite{Komineas2015} and antiskyrmions~\cite{Koshibae2016,Camosi2017,Everschor-Sitte2016,Hoffmann2017} where the winding number is opposite in sign compared to the skyrmion.

Following the pioneering work by Schryer and Walker,~\cite{Schryer1974} effective descriptions for current-driven and field-driven domain wall motion have been considered widely,~\cite{Yamanouchi2004, Thiaville2005,Tretiakov2008} including recent models leveraging canonically conjugated variables derived from the spin Berry phase action.~\cite{Berry1984a, Rodrigues2017, Ghosh2017}
For a magnetic skyrmion, it has been shown~\cite{Papanicolaou1991,Zang2011, Komineas2015} that its \textit{motion} can be described with two conjugated variables describing its position.
Effective descriptions for the internal dynamics of magnetic skyrmions, however, are few in number. 
Previous works have often focused on eigenmode analysis of magnons to obtain the small amplitude internal excitations of isolated magnetic skyrmions.~\cite{Lin2014,Schutte2014,Kravchuk2017a,Garst2017}
Skyrmion breathing modes, in which the core of the spin structure grows and shrinks periodically in time, were first described theoretically by numerical simulations of skyrmion lattices~\cite{Mochizuki2012} and later found experimentally~\cite{Onose2012} in the insulator Cu$_{2}$OSeO$_{3}$ from microwave response experiments.

In this work, we derive a non-linear effective model for rotationally symmetric (anti-)skyrmion breathing modes in terms of two collective coordinates whose validity extends beyond approaches based on an eigenmode analysis, similar to an earlier effective breathing model found in the context of dynamically stabilized skyrmions.~\cite{Zhou2015}
The novelty of this work lies in the proposal of a Hamiltonian formalism that is independent of microscopic details, applicable to the circular internal modes of both skyrmions and antiskyrmions. 
We consider an experimentally relevant model for chiral thin films to study the equilibrium and breathing properties of (anti-) skyrmions in detail and we show that all material details can be collapsed into a single effective parameter. 
 In modeling the non-linear excitations above equilibrium, we describe two dynamical regimes of coherent magnetization behavior: (i) oscillation  around the local equilibrium magnetization direction, and (ii) rotational breathing mode dynamics where the local magnetization continuously rotates. By analogy, these regimes may be thought of as a pendulum which, depending on its energy, either (i) swings about its equilibrium position or (ii) fully rotates around its pivot point.

This paper is organized as follows.
In Sec.~\ref{sec:analytical} we describe how to pass from a micromagnetic dynamical description to a Hamiltonian mechanics in terms of collective coordinates. 
 In Sec.~\ref{sec:skyrmionsanalytical} we derive the Hamiltonian mechanics describing the dynamics of soft modes for (anti-)skyrmions. We first review the Poisson bracket for the translational motion of rigid textures in two dimensions and then derive its analog for circular breathing modes.
 In Sec.~\ref{sec:MicroModel} we introduce a micromagnetic model for chiral thin films with perpendicular magnetic anisotropy, construct its effective energy as a one parameter physical model and study its equilibrium properties.
Lastly, in Sec.~\ref{sec:Breathing-Modes}, as an application of the effective Hamiltonian formalism, we study the breathing modes of circularly shaped (anti-)skyrmions and make precise predictions regarding their dissipative behavior.

\section{Collective coordinates and Hamiltonian formalism} 
\label{sec:analytical}

The magnetization dynamics below the critical temperature in ferromagnetic materials is well described in a continuum approximation by the Landau-Lifshitz-Gilbert (LLG) equation,~\cite{Gilbert2004}
\begin{equation}
\dot{\vect{m}}= \frac{1}{\mathcal{J}}\vect{m}\times \frac{\delta H[\vect{m}]}{\delta \vect{m}}+\alpha \vect{m}\times \dot{\vect{m}}, \quad (\vect{m})^{2}=1,
\label{eq:LLG}
\end{equation}
where the magnetization configuration is represented by the unit vector $\vect{m}(\vect{r},t)=\vect{M}(\vect{r},t)/M_{s}$. Here $\mathcal{J}=M_{s}/\gamma_{0}$ is the angular momentum density, $\gamma_{0}$ is the gyromagnetic constant, $M_{s}$ is the constant saturation magnetization, the Hamiltonian of the system is $H\equiv H[\vect{m}]$, $\alpha$ is the Gilbert damping parameter, and the overdots indicate total time derivatives $\dot{\vect{m}}\equiv d\vect{m}/dt$.

The LLG is a non-linear differential equation with an infinite number of degrees of freedom. 
This poses an obstacle for the comprehensive study of the magnetization dynamics, prohibiting a full analytical solution and usually requiring extensive micromagnetic simulations. 
However,  the low energy dynamics of magnetic textures depend only on the system's \textit{soft modes} which can be described by conjugated variables in a Hamiltonian formalism.~\cite{Clarke2008,Tretiakov2008,Rodrigues2017,Ghosh2017} 
This allows access to generic features of the magnetization textures independently from the microscopic characteristics of the material. 

The conservative, precessional, part of the LLG equation may be derived from the first variation of the action 
\begin{equation}
\mathcal{S}=\int dt L = \mathcal{S}_{B}-\int dt H,
\label{eq:Micromagnetic-action}
\end{equation}
with the constraint of a constant magnetization amplitude. Here $L$ is the Lagrangian and $\mathcal{S}_{B}$ is the spin Berry phase action.~\cite{Berry1984a,Kosevich1990,Papanicolaou1991} 
The spin Berry phase couples the dynamical degrees of freedom of the local magnetization. 
In a pherical representation of the magnetization field, $\vect{m}(\theta(\vect{r},t),\phi(\vect{r},t )) = (\sin\theta\cos\phi,\sin\theta\sin\phi,\cos\theta)^{\intercal}$, using the ``north-pole'' parametrization,~\cite{Braun1996} it is possible to write the spin Berry phase as,
\begin{equation}
\mathcal{S}_{B} =\mathcal{J} \int dt \int dV (1-\cos\theta)\dot{\phi}.
\label{eq:Spin-BP-action}
\end{equation} 

Two canonically conjugated fields, in this case given by $\phi(\vect{r},t)$ and $\mathcal{J} [1-\cos(\theta(\vect{r},t))]$,
are sufficient to describe the magnetization dynamics. 
 In addition to the energy conserving part, the phenomenological damping term 
 may be introduced directly into the Euler-Lagrange equations by use of a Rayleigh dissipative functional,~\cite{Gilbert2004} such that the full Eq.~\eqref{eq:LLG} is derived from 
 \begin{equation}
0 = \frac{d}{dt}\frac{\delta L[\vect{M},\dot{\vect{M}}]}{\delta \dot{\vect{M}}}-\frac{\delta L[\vect{M},\dot{\vect{M}}]}{\delta \vect{M}}+\frac{\delta \mathcal{R}[\dot{\vect{M}}]}{\delta \dot{\vect{M}}},
\label{eq:Euler-Lagrange}
\end{equation}
where the Rayleigh functional is
\begin{equation}
\mathcal{R}[\dot{\vect{M}}]=\frac{\alpha\mathcal{J}}{2M_{s}^{2}} \int dV (\dot{\vect{M}})^{2}.
\label{eq:Rayleigh-functional}
 \end{equation}
 
Since the magnetization is characterized by a field, it has infinitely many modes. 
It is possible to map the dynamics to an infinite number of time dependent functions $\vect{\xi}(t)= \{ \xi_1(t), \xi_{2}(t), \dots\}$, i.e.\ $\vect{m}[\vect{r},\vect{\xi}(t)]$. 
The unique equation of motion with infinite degrees of freedom, Eq.~\eqref{eq:LLG}, becomes an infinite set of equations of motion for these dynamical parameters in this approach. 
The utility of this mapping is that these different parameters may have different time scales. 
Therefore the low energy excitations can be described as a reduced set of collective coordinates, $\vect{\xi}(t)= \{ \xi_1(t), \xi_{2}(t), \dots, \xi_{2N}(t)\}$ describing the soft modes which dominate the dynamics,~\cite{Tretiakov2008} and whose relaxation time is much longer than the rest of the infinite set. 
Examples include the position and tilt angle of a domain wall when subjected to small driving currents, or the position $(X,Y)^{\intercal}$ of a rigid homogenous domain in steady translational motion as described originally by Thiele.~\cite{Thiele1973} 
Identifying collective coordinates therefore offers the possibility to work with a reduced number of degrees of freedom, i.e.\ a minimal number of equations of motion just for the corresponding soft modes, rather than the full LLG field equation, which is usually analytically intractable.

There are several methods to obtain the equations of motion for the collective coordinates.~\cite{Thiele1973,Tretiakov2008,Everschor2011,Rodrigues2017} 
In this paper we outline the Hamiltonian approach. 
Within this formalism, the equations of motion are obtained in an explicit and direct manner from an effective energy which is a function of collective coordinates for the soft modes. 
We provide the comparison between this Hamiltonian approach and the generalized Thiele approach in Appendix~\ref{app:hamiltonian-thiele-comparison}.
Recasting Eq.~\eqref{eq:LLG} in a Hamiltonian language yields~\cite{Papanicolaou1991} 
\begin{equation}
\dot{\vect{m}}=\{ \vect{m}, H \}_{\Phi,\Pi} + \gamma_{\vect{m}},
\label{eq:LLG-Hamiltonian-form}
\end{equation}
containing an energy conserving part expressed using Poisson brackets for two canonically conjugated fields $\Phi(\vect{r},t)$ and $\Pi(\vect{r},t)$,
and a damping contribution described by $\gamma_{\vect{m}}\equiv \alpha \vect{m}\times \dot{\vect{m}}$. 
The conventions for Poisson brackets used throughout this paper and their properties are listed in Appendix~\ref{app:PBs}.

The effective equations of motion for the collective coordinates in the Hamiltonian language are
\begin{equation}
\dot{\xi}_{i} = \{\xi_{i},H\}_{\vect{q},\vect{p}} + \gamma_{\xi_{i}},
\label{eq:xi-dot}
\end{equation}
where the Poisson brackets $\{\cdot,\cdot\}_{\vect{q},\vect{p}}$ are now defined for pairs of canonical variables ($q_{i},p_{i}$) in terms of the independent collective coordinates $\xi_{i}(t)$ and their time derivatives $\dot{\xi}_{i}(t)$.

In general, the canonical coordinates may always be taken as collective coordinates themselves, $q_{i}\equiv\xi_{i}$; meanwhile, the momenta are gauge-dependent functionals $p_{i}=A_{i}[\vect{\xi}]$ so that the corresponding Lagrangian is
$L=\sum_{i}A_{i}\dot{\xi}_{i}-H$.~\cite{Zakharov1997}
The canonical momenta arising from the vector potential correspond to a monopole field in the spin Berry phase action, i.e.\ $\nabla \times \vect A = \mathcal{J}\vect m$,~\cite{Clarke2008,Tchernyshyov2015}
and are related to the gyrotropic tensor by~\cite{Tretiakov2008,Clarke2008}
$\mathcal{G}_{ij}=\mathcal{J}^{-1}(\partial A_{i}/\partial\xi_{j} -  \partial A_{j}/\partial \xi_{i} )$.
For the commonly-used gauge choice giving Eq.~\eqref{eq:Spin-BP-action} the functional is simply $A_{i}[\vect{\xi}]=\int dV (1-\cos\theta)\partial_{\xi_{i}}\phi$. 
In terms of the magnetization, the gyrotropic tensor is
\begin{equation}\label{eq:Giro-tensor}
\mathcal{G}_{ij}=\int dV \vect{m}\cdot \left(\partial_{\xi_{i}}\vect{m}\times\partial_{\xi_{j}}\vect{m} \right).
\end{equation}

In the following, we will only consider Hamiltonians $H=H[\vect{\xi}]$ that are not explicitly time dependent. 
In this case the dissipative part $\gamma_{\xi_{i}}$ is derived by comparing the rate of energy dissipation that is given by the dissipative functional, $\dot{H}=-2\mathcal{R}=-\alpha\mathcal{J}\sum_{ij}\mathcal{D}_{ij}\dot{\xi}_{i}\dot{\xi}_{j}$, to its expansion $\dot{H}=\sum_{i}\frac{\partial H}{\partial \xi_{i}}\dot{\xi}_{i}$, where 
\begin{equation}\label{eq:Visco-tensor}
\mathcal{D}_{ij}= \int dV \left(\partial_{\xi_{i}}\vect{m}\cdot\partial_{\xi_{j}}\vect{m}\right)
\end{equation}
is the viscosity tensor. 
Solving for $\gamma_{\xi_{i}}$ yields the result,
\begin{equation}
\gamma_{\xi_{i}}=\alpha \mathcal{J}\sum_{j,k}\{\xi_{i},\xi_{j} \}_{\vect{\xi},\vect{p}_{\vect{\xi}}}\mathcal{D}_{jk}\dot{\xi}_{k}.
\label{eq:gamma-general-result}
\end{equation}

We point out that equations~\eqref{eq:xi-dot} and~\eqref{eq:gamma-general-result} taken together are not specific to magnetization dynamics; rather, they are generic for conservative mechanical systems to which frictional forces (force terms linear in velocities) are included in the equation of motion by the use of a Rayleigh function.
Using the identity $\{\xi_{i},H\}_{\vect{\xi},\vect{p}_{\vect{\xi}}}=\sum_{j}\{\xi_{i},\xi_{j}\}_{\vect{\xi},\vect{p}_{\vect{\xi}}}\frac{\partial H}{\partial\xi_{j}}$, we may write  Eqs.~\eqref{eq:xi-dot} and \eqref{eq:gamma-general-result} together for a reduced set of soft modes,
\begin{align}
\dot{\xi}_{i} &= \sum_{j=1}^{2N}\{\xi_{i},\xi_{j}\}_{\vect{\xi},\vect{p}_{\vect{\xi}}}\left(\frac{\partial H}{\partial \xi_{j}} + \alpha \mathcal{J}\sum_{k}\mathcal{D}_{jk}\dot{\xi}_{k}\right), 
\label{eq:xi-dot-full}
\end{align}
which is our first main theoretical result. 
As an explicit matrix equation the above result becomes 
\begin{equation}
\vect{\dot{\xi}}=\frac{1}{\mathcal{J}}(\mathcal{G}-\alpha\mathcal{D})^{-1} \frac{\partial H}{\partial \vect{\xi}} .\label{eq:ours-explicit}
\end{equation}
by use of the relation between the Poisson brackets and the gyrocoupling tensor\cite{Tchernyshyov2015} $ \mathcal{J} \{\xi_{i},\xi_{j} \}_{\vect{\xi},\vect{p}_{\vect{\xi}}}=(\mathcal{G}^{-1})_{ij}$.

\section{Effective Hamiltonian descriptions for magnetic skyrmions} 
\label{sec:skyrmionsanalytical}

The LLG equation contains topologically non-trivial solutions. 
In 1D they include domain walls, and in 2D they include different types of solitons that are distinguished by their integer topological charge or winding number,
\begin{align}
    \mathcal{Q} = \frac{1}{4\pi}\int dxdy\, \vect{m}\cdot\left(\partial_{x}\vect{m}\times\partial_{y}\vect{m}\right). \label{eq:topcharge}
\end{align}
An important example of magnetic solitons in 2D are skyrmions. 
In the following we will apply the Hamiltonian description given by Eqs.~\eqref{eq:xi-dot} and~\eqref{eq:gamma-general-result} to the dynamics of skyrmions and antiskyrmions.
First we review the steady translational motion for rigid topological structures using this approach, and second we apply the technique to study the breathing mode.

\subsection{Translational modes of rigid topological textures} 
\label{sec:translation-modes}

The translational motion of a rigid structure can be described in terms of a position $\vect{r}_{s}(t) =(X(t),Y(t))^{\intercal}$ where $X$ and $Y$ are collective coordinates describing a soft mode.\cite{Papanicolaou1991, Moutafis2009, Tchernyshyov2015} 
The following discussion does not require a specific definition of $\vect{r}_s(t)$.
To obtain the Poisson bracket structure from Eq.~\eqref{eq:Spin-BP-action}, we need to peturbatively expand the fields $\theta(\vect{r},t)$ and $\partial_{t}{\phi(\vect{r},t)}$ in terms of small deviations in $X$,$Y$ and $\dot{X},\dot{Y}$ respectively. Considering a rigid texture ansatz for the magnetization, $\vect{m}(\vect{r},t)=\vect{m}_{0}(\vect{r}-\vect{r_{s}}(t))$, implies~\cite{Thiele1973} $\partial_{X}\vect{m}=-\partial_{x}\vect{m}$ and $\partial_{Y}\vect{m}=-\partial_{y}\vect{m}$.
Performing an expansion in the spin Berry Phase action~\eqref{eq:Spin-BP-action} up to quadratic order and discarding terms that do not contribute to the dynamics leads to, (see Appendix \ref{app:XY-PB-derivation})
\begin{align}
\mathcal{S}_{B} \approx \int dt  z_{0} \mathcal{J} X\dot{Y} \int dxdy\, \left[(\partial_{y}\cos\theta) \partial_{x}\phi - (x\leftrightarrow y)\right],
\label{eq:SbXY}
\end{align}
where $z_{0}$ is the thickness of the system and the spatial integral is proportional to the topological charge defined in Eq.~\eqref{eq:topcharge}. 
Hence the effective action is
\begin{equation}
S_{\text{eff}}=\int dt (4\pi \mathcal{Q} z_{0} \mathcal{J} X\dot{Y} - H).
\end{equation}
Noting that the canonical momentum to $Y$ is $p_{Y}= 4\pi \mathcal{Q} z_{0}\mathcal{J}X$, we therefore read off the Poisson bracket for topologically non trivial textures,~\cite{Papanicolaou1991} 
\begin{align}
\{Y, X \}_{Y,p_{Y}}=\frac{1}{4\pi\mathcal{Q} z_{0}\mathcal{J}}.
\label{eq:XY-PB}
\end{align}
Inserting the Poisson bracket of Eq.~\eqref{eq:XY-PB} into Eq.~\eqref{eq:xi-dot-full} gives the dynamical equations of motion for the translational mode
\begin{subequations}
\label{eq:dotXY}
\begin{align}
 \dot{X} &= -\frac{1}{4\pi  \mathcal{Q} z_{0}}\left[\frac{1}{\mathcal{J}}\frac{\partial H}{\partial Y} +\alpha (\mathcal{D}_{YX}\dot{X} + \mathcal{D}_{YY}\dot{Y} )\right] \\
\dot{Y}&=\frac{1}{4\pi  \mathcal{Q} z_{0}}\left[\frac{1}{\mathcal{J}}\frac{\partial H}{\partial X} +\alpha(\mathcal{D}_{XX}\dot{X} +  \mathcal{D}_{XY}\dot{Y})\right],
\end{align}
\end{subequations}
which are equivalent to the traditional Thiele equations for skyrmions,~\cite{Thiele1973,Everschor2012a} and in the case of circular states, like for a simple skyrmion, the off-diagonal elements $D_{XY}=D_{YX}$ vanish.

\begin{figure}[t]
    \includegraphics[width=\columnwidth]{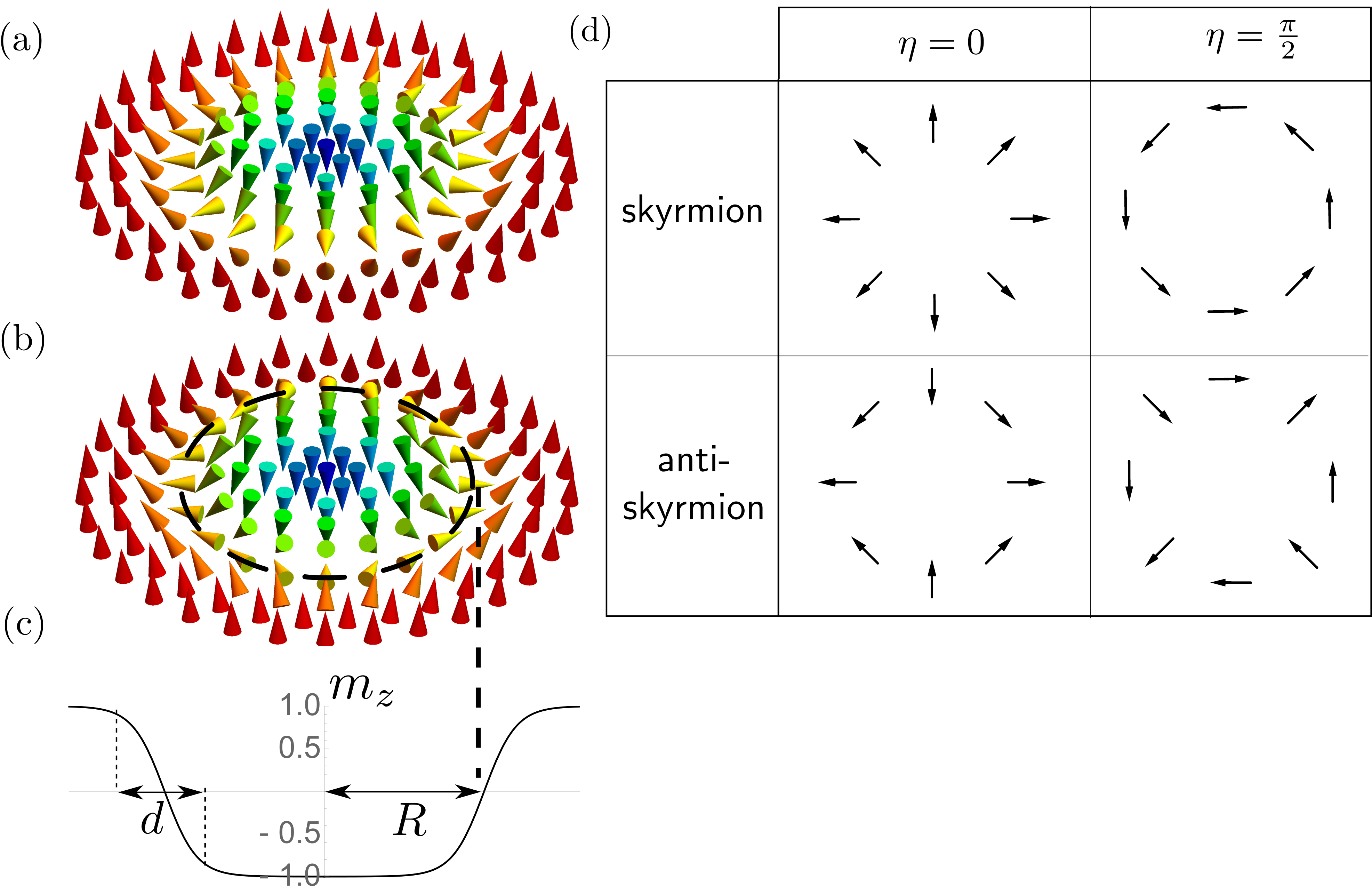}
    \caption{
    Schematic figure of a (a) N\'{e}el skyrmion (b) antiskyrmion and (c) the common $m_{z}$ component for both structures. 
    (d) The in-plane spins for prototypical (anti-)skyrmions with phase $\eta$, where $\eta=0$ corresponds to the spin structures to the left in (a) and (b).
  }  
    \label{fig:schematic-Neel-idea}
\end{figure}

\subsection{Internal dynamics of skyrmions and antiskyrmions} 
\label{sec:internal-circular-dynamics}

For a study of the internal dynamics of magnetic solitons, such as skyrmion breathing modes, one must go beyond the rigid texture (or traveling-wave) approximation used in the last section.  
We will consider a thin film system that is translationally invariant along the $z$ direction and contains localized rotationally symmetric states,\cite{Bogdanov1989, Bogdanov1994, Nagaosa2013}
\begin{subequations}
 \label{eq:mag-ansatz}
\begin{align}
&\vect{m}(\vect{\rho}) = \left(\sin\theta(\rho)\cos\phi(\psi),\sin\theta(\rho)\sin\phi(\psi),\cos\theta(\rho)\right)^{\intercal}, \\
&\vect{\rho}=(\rho\cos\psi,\rho\sin\psi)^{\intercal},
\end{align} 
\end{subequations}
where $\rho$ and $\psi$ are the polar coordinates in the $x$--$y$ plane.
Since the magnetic angle $\theta$ only changes with distance $\rho$ from the skyrmion core, these field configurations are referred to as \emph{circular} in this work.
In view of describing the breathing modes of skyrmions as well as antiskyrmions we parameterize the azimuthal angle of the magnetization by,
\begin{equation}
\phi(\psi) = m\psi + \eta,
\label{eq:phi-ansatz}
\end{equation}
where $m\in \mathbb{Z}$ is the vorticity and $\eta$ is the relative azimuthal angle.
For a simple skyrmion one has $m=1$ and the angle $\eta$ describes its helicity: for a Bloch-skyrmion $\eta=(n+1/2)\pi$ and for a N\'{e}el-skyrmion $\eta = n\pi$ where $n\in \mathbb{Z}$, see Fig.~\ref{fig:schematic-Neel-idea}. 
This corresponds to $\mathcal{Q}= -m/2 \left[\cos\theta(\infty) - \cos\theta(0)\right]$ with the boundary conditions  $\theta(0)=\pi$ and $\theta(\infty) = 0$. 
In other words the ferromagnetic background points in the $+\hat{\vect{z}}$ direction.
For circular (anti-)skyrmions there are two characteristic length scales: the radius $R$ and the \textit{width} $d$ for the twisted domain over which $m_{z}$ varies, see Fig.~\ref{fig:schematic-Neel-idea}(c). 
We define the radius $R$ by the circle where $m_{z}=0$. 
We assume in the following that during the breathing dynamics $\theta(\rho)$ retains its smooth and monotonic variation, ensuring the definition of $R$ to be unique. 
This work will consider only large (anti-)skyrmions in the regime where  $d \ll R$ where the skyrmion's wall width can be considered constant even as its radius is allowed to vary.

The relevant soft mode for the breathing dynamics of a circular (anti-)skyrmion is described by the radius $R(t)$ and the relative azimuthal angle $\eta(t)$.~\cite{Zhou2015} 
Unlike in Sec.~\ref{sec:translation-modes}, the magnetic texture is not rigid but soft in its overall shape. 
For skyrmions, the phase $\eta$ is the global in-plane angle of the local magnetization pointing away from the radial direction (such that $\eta=0,\pi$ correspond to N\'eel and $\eta=\pm\pi/2$ to Bloch configurations respectively), while for antiskyrmions, changing $\eta$ corresponds to a rigid rotation of the entire magnetic texture.

We will now follow the general recipe of Sec.~\ref{sec:analytical}. 
By estimating the volume integral for the Berry phase term in Eq.~\eqref{eq:Spin-BP-action} we obtain the effective action 
\begin{equation}
\mathcal{S}_{\text{eff}}= \int dt \left(a\mathcal{J} R^{2}\dot\eta - H\right),
\label{eq:effective-S}
\end{equation}
in terms of the softmodes $R$ and $\eta$. The length scale $a$ is given by $a=2\pi z_{0} C$, where $C$ is a  dimensionless constant (with $0<C<2$) arising from the integral over $\rho$. 
Defining the canonical momentum conjugate to $\eta$ as $p_{\eta}= a \mathcal{J} R^2$, we read off the Poisson bracket as 
\begin{equation}
\label{eq:Reta}
\{\eta,  R^2\}_{\eta,p_{\eta}} = \frac{1}{\mathcal{J} a}.
\end{equation}
Via the identity: $\{\eta,R^{2}\}_{\eta,p_{\eta}}=(\partial_{R}R^{2})\{\eta,R\}_{\eta,p_{\eta}}=2R\{\eta,R\}_{\eta,p_{\eta}}$ we obtain $\{\eta,R\}_{\eta,p_{\eta}}=(2a\mathcal{J}R)^{-1}$. Exploiting Eq.~\eqref{eq:xi-dot-full}, the dynamical equations for $\eta$ and $R$ are readily derived
\begin{subequations}
\label{eq:dotReta}
\begin{align}
 \dot{\eta} &= \frac{1}{2 a \mathcal{J} R}\frac{\partial H}{\partial R} +\alpha \frac{\mathcal{D}_{RR}}{2a}\frac{\dot{R}}{R}, \label{eq:doteta}\\
\dot{R}&=-\frac{1}{2 a \mathcal{J} R}\frac{\partial H}{\partial \eta} -\alpha\frac{\mathcal{D}_{\eta\eta}}{2a}\frac{\dot\eta }{R}. \label{eq:dotR}
\end{align}
\end{subequations}
Above we used that $\partial_{R}\vect{m}\cdot \partial_{\eta}\vect{m}=0$ by virtue of the rotationally symmetric ansatz Eqs.~\eqref{eq:mag-ansatz} and~\eqref{eq:phi-ansatz}. 
Eq.~\eqref{eq:dotReta} describes the effective internal dynamics of a rotationally symmetric magnetic texture subject to the ansatz Eqs.~\eqref{eq:mag-ansatz} and~\eqref{eq:phi-ansatz} for time-independent Hamiltonians $H$. 

\section{Effective energy for circular skyrmions}
\label{sec:MicroModel}

Previous works have assumed an explicit domain wall ansatz for the skyrmion's radial profile.~\cite{Sheka2001,Rohart2013,Romming2015} 
By using scaling arguments, however, one does not need to assume a specific ansatz for the skyrmion. 
The energy can in fact be expanded in powers of the collective $R$ coordinate for skyrmions satisfying $d\ll R$. 
We illustrate this procedure for a micromagnetic model including exchange, anisotropy and interfacial Dzyaloshinskii-Moriya interaction (DMI), whose magnetic free energy is given by
\begin{align}
H [ \vect{m} ] = \int d V \{ & A \sum _ { i } \left( \partial _ { i } \vect{m} \right) ^ { 2} + K \left( 1- m _ { z } ^ { 2} \right)  \notag \\
&+ D(m_{z}\partial_{x}m_{x}-m_{x}\partial_{x}m_{z}) \notag \\
&\pm D(m_{z}\partial_{y}m_{y}-m_{y}\partial_{y}m_{z}) \},\label{eq:efunctional}
\end{align}
where the + (-) sign stands for the isotropic (anisotropic) DMI which stabilizes circular (anti-)skyrmions. Performing the expansion of Eq.~\eqref{eq:efunctional} in $R$ for a large radius skyrmion we obtain the following dimensionless effective energy in units of $\mathcal{E}_{\mathrm{DW}}=A^{3/2}K^{-1/2}$,
\begin{equation}
\tilde{E}(\tilde{r},\eta) = (c_{1} - c_{3} g \cos\eta) \tilde{r} + \frac{c_{2}}{\tilde{r}},\label{eq:effectiveE}
\end{equation}
where $\tilde{r}\equiv R/\Delta$ is the dimensionless (anti-)skyrmion radius in units of the one-dimensional domain wall width $\Delta = \sqrt{A/K}$.
The single coupling constant $g$ is the reduced DMI strength $g = \pi D/(4\sqrt{AK})$ selecting for either a ferromagnetic $|g|<1$ or helical $|g|>1$ mean-field ground state.~\cite{Bogdanov1989,Bogdanov1994,Rohart2013} 
All rescaling constants are summarized in Table~\ref{tab:units}. 
The dimensionless values $c_1, c_2$ and $c_3$ are uniquely determined by the material parameters through the coupling constant $g$ (see Appendix \ref{app:ExtraSecIV}),~\cite{Kravchuk2017a} since we have chosen to focus on the limit where the skyrmion wall width $d$ is time-independent. 
The effective energy is identical for both skyrmions and antiskyrmions as a consequence of Eq.~\eqref{eq:phi-ansatz} taken with the appropriate choice of $m=+1 (-1)$ for skyrmions (antiskyrmions). 
As such, we will reduce our discussion to skyrmions only from now on even though the results derived are valid for antiskyrmions as well. In the Appendix \ref{app:ExtraSecIV}, we discuss how to introduce other interactions, such as dipole-dipole and bulk DMI into this framework.

\subsection{Energy landscape analysis}
\label{sec:Static-investigations}

\begin{figure}[t]
	\centering
	\includegraphics[width=\columnwidth]{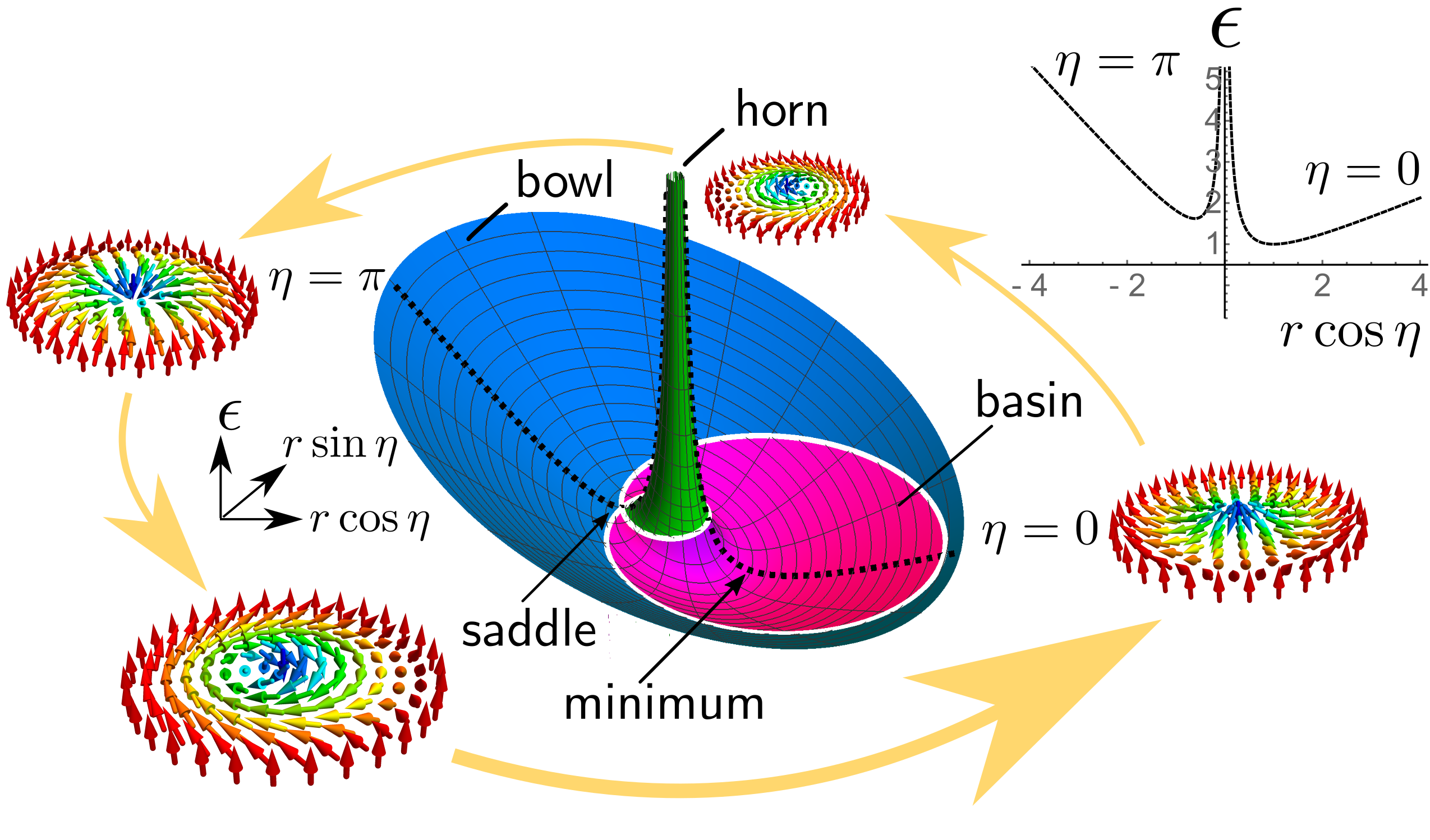}
		 \caption{
		 Effective energy landscape for circular skyrmion breathing modes, Eq.~\eqref{eq:rescaled-energy}, with $B=0.45$, highlighting the distinct {\it basin}, {\it bowl} and {\it horn} energy partitioned by the constant saddle energy curve (shown in white). 
		 Inset: a cut through $\eta=0$ and $\eta=\pi$. 
		 }
		\label{fig:energy-landscape-schematic}
\end{figure}

\begin{table}[t]
\begin{center}
$
\begin{array}{l| l|c|l}  \hline
\text{Quantity} & \text{Characteristic} & \text{Definition} & \text{SI unit} \\    [\medskipamount] \hline
\Delta & \text{length}  &  A^{1/2}K^{-1/2} & \text{m} \\ 
1/\omega_{\mathrm{FM}} & \text{time}& M_{s}\gamma_{0}^{-1} K^{-1} & \text{s}  \\
\mathcal{E}_{DW} & \text{energy}& A^{3/2}K^{-1/2} & \text{J} \\ 
g &\text{coupling constant}& \pi D (4\sqrt{AK})^{-1} & 1 \\ [\medskipamount] \hline
\end{array}
$
\end{center}
\caption{Natural unit system for a study of skyrmion breathing modes in a system with DMI, exchange and perpendicular magnetic anisotropy.}
\label{tab:units}
\end{table}

The energy landscape described by the effective model~\eqref{eq:effectiveE} has two extrema. 
The first is a global energy minimum with corresponding equilibrium coordinates
\begin{equation}
\label{eq:ReqDefinition}
\tilde{r}_{\mathrm{eq}} = \sqrt{\frac{c_{2}}{c_{1} - c_{3}|g|}}, \quad \eta_{\mathrm{eq}}=\left\{\begin{array}{lll}
0 & \text{if} & 0<g<1 \\
\pi & \text{if} &-1<g<0.
\end{array}\right. 
\end{equation}
The second represents an energy saddle point with coordinates:
\begin{equation}
\tilde{r}_{\mathrm{sad}} = \sqrt{\frac{c_{2}}{c_{1} + c_{3}|g|}}, \quad \eta_{\mathrm{sad}}=\left\{\begin{array}{lll}
\pi & \text{if} &0<g<1 \\
0 & \text{if} &-1<g<0,
\end{array}\right. 
\end{equation}
where we note that $\tilde{r}_{\mathrm{eq}}>\tilde{r}_{\mathrm{sad}}$ and that their respective effective energies are $\mathrm{E}_{\mathrm{eq(sad)}}=2c_2/\tilde{r}_{\mathrm{eq(sad)}}$.

Upon rescaling the radii by the equilibrium radius ($r\equiv \tilde{r}/\tilde{r}_{\mathrm{eq}}$) and energy by the equilibrium energy ($\epsilon\equiv E/E_{\mathrm{eq}}$) in \eqref{eq:effectiveE}, the effective energy reduces to the simplified form
\begin{equation}
\epsilon  = \frac{1}{2r}\left(\frac{1-B\cos\eta}{1-|B|}r^{2}+1 \right),
\label{eq:rescaled-energy}
\end{equation}
where we have defined $B=c_{3}g/c_{1}$ as the single parameter which encapsulates the entire contribution from the material properties on the physics of the system. 
In these units the saddle point radius is exactly the inverse of the corresponding saddle point energy ($r_{\mathrm{sad}}= \epsilon_{\mathrm{sad}}^{-1} = \sqrt{(1-|B|)/(1+|B|)}$).

\begin{figure}[t]
    \includegraphics[width=\columnwidth]{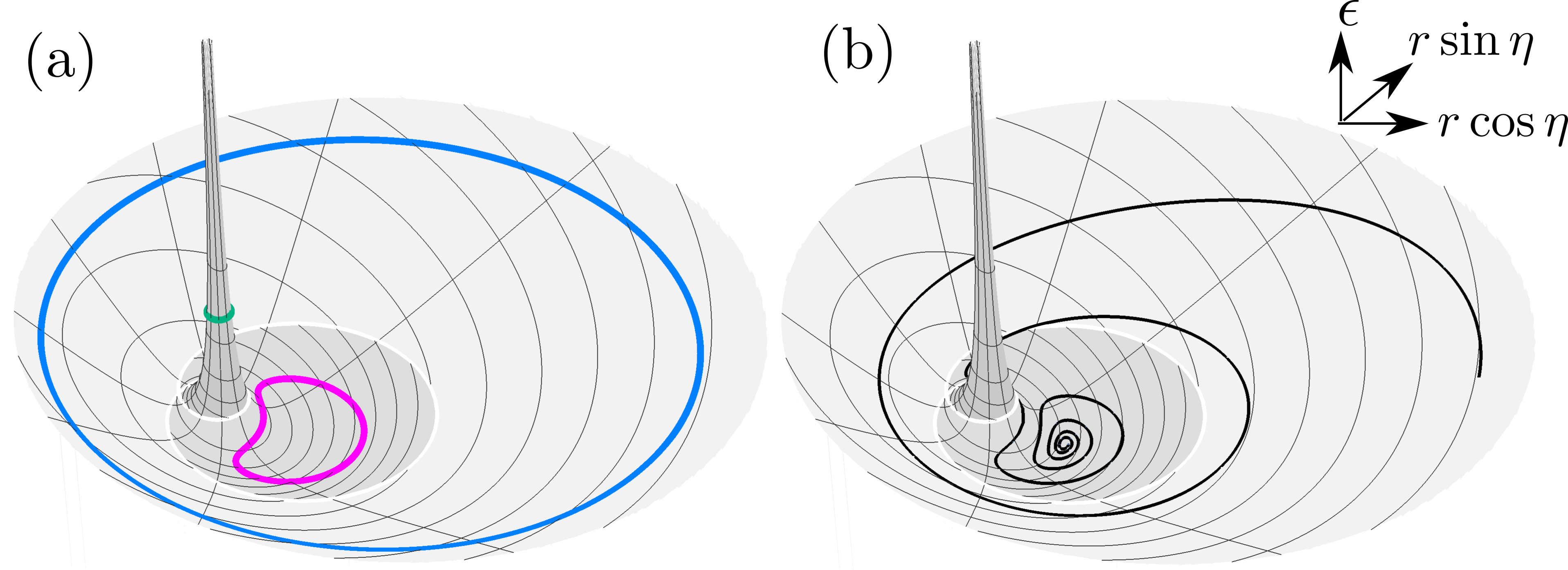}
    \caption{
    Breathing mode dynamics from the energy landscape with $B=0.45$. 
    (a) Constant energy orbits: two are degenerate on the bowl and horn at $\epsilon=2\epsilon_{\mathrm{sad}}$, while the third one lies in the basin at $\epsilon=\frac{3}{4}\epsilon_{\mathrm{sad}}$. 
    (b) A damped trajectory starting on the bowl that spirals in towards the energy minimum.
    }
    \label{fig:e-landscape}
\end{figure}

A cut along the $\epsilon=\epsilon_{\mathrm{sad}}$ plane partitions the energy landscape into three distinct regions. 
A schematic view of the energy landscape is shown in Fig.~\ref{fig:energy-landscape-schematic} where individual energy sectors have been color coded to guide the reader.
The $(r,\eta)$ coordinates in the {\it bowl} and {\it horn} regions both correspond to high energy states ($\epsilon>\epsilon_{\mathrm{sad}}$) as opposed to the {\it basin}'s low energy states ($\epsilon<\epsilon_{\mathrm{sad}}$). 
In the inset of Fig.~\ref{fig:energy-landscape-schematic} we show a cut through $\eta=0$ and $\eta=\pi$ to emphasize the structure of the extrema introduced above. 
The constant energy trajectories followed by skyrmions in their configuration space in the absence of damping (see Fig.~\ref{fig:e-landscape}(a)) are obtained by solving Eq.~\eqref{eq:rescaled-energy} for the rescaled radius
\begin{equation}
\label{eq:const-E-orbit}
r_{\pm}(\eta) = \frac{(1-|B|)\epsilon}{1-B\cos\eta}\left[1\pm\sqrt{ 1-\frac{1-B\cos\eta}{(1-|B|)\epsilon^2} }\,\right].
\end{equation}
While the two solutions in \eqref{eq:const-E-orbit} represent distinct {\it horn} and {\it bowl} orbits in the $\epsilon>\epsilon_{\mathrm{sad}}$ regime, they represent the two branches of the same {\it basin} orbit in the $\epsilon<\epsilon_{\mathrm{sad}}$ case. 
In all scenarios, constant energy orbits describe skyrmion breathing motions as the radius grows and shrinks as a function of $\eta$. 
The qualitative nature of orbits in the {\it basin} and {\it bowl/horn} regions are however very different from each other as the dynamical range of $\eta$ is limited in the basin orbits while it takes all values from $0$ to $2\pi$ in the horn/bowl orbits. 
This leads us to denominate the high and low energy breathing dynamics as {\bf rotating} and {\bf oscillating} modes respectively. 
The degeneracy of the rotating modes disappears as their energy is lowered through the saddle energy and into the basin. 
The horn rotations are unphysical however as they would predict uncollapsable skyrmions in the limit of very small skyrmion sizes. Furthermore, since our theory is only applicable for skyrmion much larger than the profile wall width $d$, it cannot reliably describe their behavior at such small radii.
Overall, the dynamical spectrum of this model is reminiscent of that of a simple pendulum which exhibits rotations and oscillations around its suspension point depending on whether the kinetic energy is greater or less than the potential energy of its ``upside down'' unstable equilibrium.

Lastly, the full range of the radial oscillations can be readily obtained from Eq.~\eqref{eq:const-E-orbit} by noting that all radial maxima/minima in the orbits appear on the $\eta=0,\pi$ line (as shown in Fig.~\ref{fig:energy-landscape-schematic}). 
For $B>0$ the rotating modes have $r_{\mathrm{max}}=r_{\pm}(0)$ and $r_{\mathrm{min}}=r_{\pm}(\pi)$ whereas in the oscillating phase one has $r_{\mathrm{max}}=r_{+}(0)$ and $r_{\mathrm{min}}=r_{-}(0)$. 
Since the lower energy branch on this line is independent of the coupling $B$, the amplitude for oscillating breathing modes below the saddle energy is insensitive to material properties (see discussion below).

\section{Skyrmion breathing modes}
\label{sec:Breathing-Modes}

\subsection{Equations of Motion}
\label{sec:breathing-modes}

\begin{figure*}[t]
    \includegraphics[width=\textwidth]{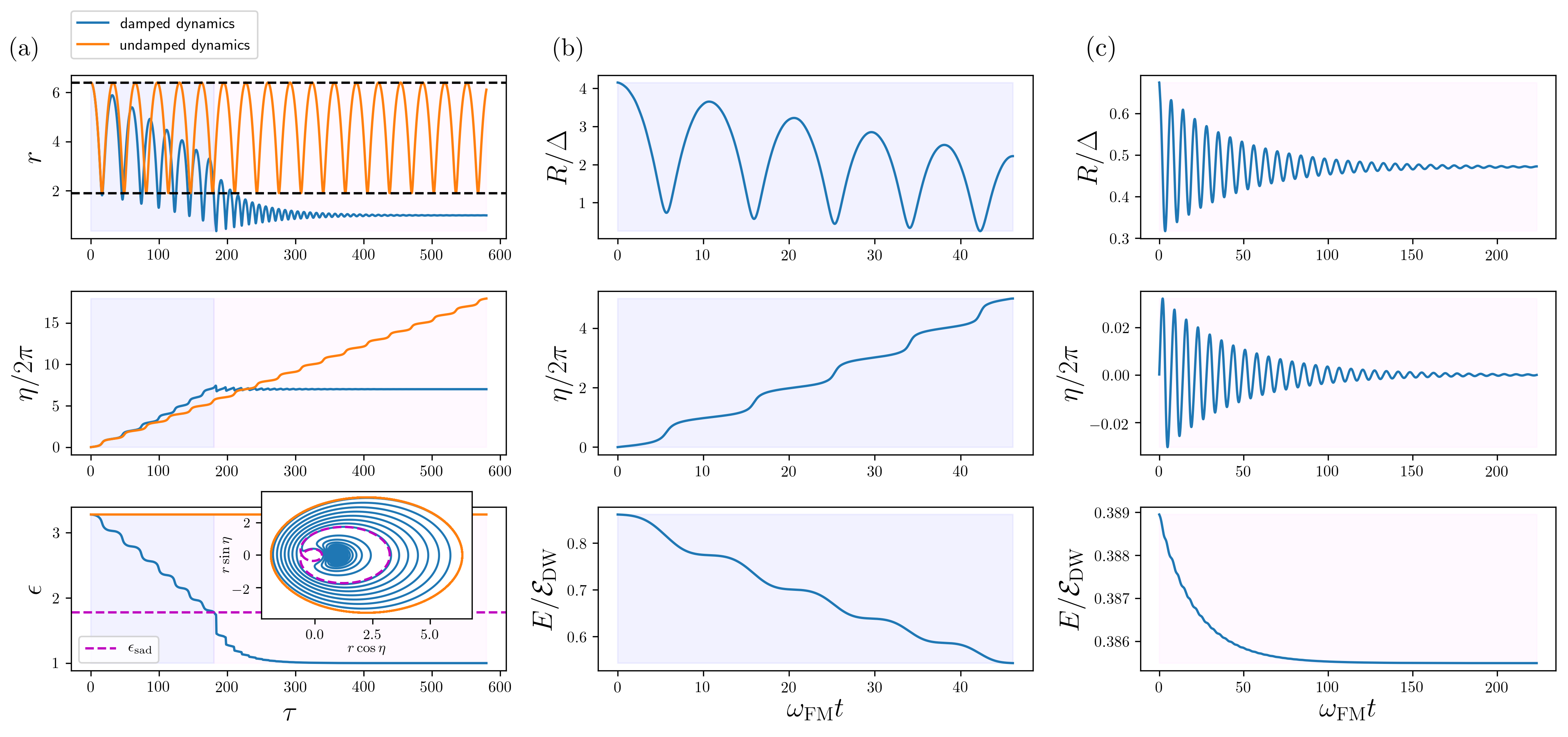}
    \caption{
    Time evolution of the dynamical system for the rescaled radius $r$ (top row), azimuthal angle $\eta$ (middle row), and rescaled energy (bottom row).
    Orange and blue curves correspond to undamped and damped trajectories respectively. 
    (a) Numerical results based on the effective model, Eq.~\eqref{eq:reparametrised-system} with $B=0.45$ for an initial energy starting in the $\epsilon>\epsilon_{\mathrm{sad}}$ bowl region (light blue background).
    The black dashed lines in (a) denote the maximum and minimum radii of the undamped trajectory ($r_{1}^{+}$ and $r_{2}^{+}$ from Eq.~\eqref{eq:max-min-radiuses} respectively). 
    The transition from rotating to oscillating phase (light pink background) is best seen in the behavior of the azimuthal angle where $\eta$ rotates a number of times before oscillating around its equilibrium value. 
    Inset: projection of the trajectory on the $(r\cos\eta, r\sin\eta)$ polar coordinate plane where the dashed magenta line represents saddle energy $\epsilon_{\mathrm{sad}}$ separatrix. 
    Subplots (b) and (c) show the breathing behavior of skyrmions obtained by micromagnetic simulations for an energy above and below the saddle point energy respectively. 
    The parameters of the simulations are $\alpha=0.02$, $D=2.8\times 10^{-3}$ Jm$^{-2}$, $A=1.5\times 10^{-11}$ Jm$^{-1}$, $K=1.1\times 10^{6}$ Jm$^{-3}$, i.e.\ $g\approx 0.54$. 
    The qualitatively different energy decay behavior is shown in in the lower row to transition from linear-like to exponentially decreasing in the rotating and oscillating phases respectively.
    }
    \label{fig:high-energy-oscillation}
\end{figure*}

In terms of the variables $r$ and $\eta$, the Poisson bracket Eq.~\eqref{eq:Reta} becomes
\begin{equation}
\{\eta, r^{2}\}_{\eta,\tilde{p}_{\eta}} = c_{0}/ \tilde{r}_{\mathrm{eq}}^{2}, \label{eq:PB-rescaled}
\end{equation}
where we assume a \textit{constant} shape factor $c_{0}$ coming from the dimensionless integral $c_{0}^{-1}=2\pi \tilde{z}_{0}\int_{0}^{\infty}dx\,x\, [1-\cos\theta(Rx)]$ in the spin Berry phase action. 
Moreover the dimensionless dissipation factors both scale linearly with the radius $\tilde{r}$,
\begin{align}
\label{eq:treatment-of-damping}
\mathcal{D}_{rr} = c_{r}\tilde{r}=c_{r}\tilde{r}_{\mathrm{eq}}r, \quad \mathcal{D}_{\eta\eta}=c_{\eta}\tilde{r}=c_{\eta}\tilde{r}_{\mathrm{eq}}r
\end{align}
where $c_{r}$ and $c_{\eta}$ are time-independent proportionality constants that may depend on the coupling strength $B$.
Using Eqs.~\eqref{eq:PB-rescaled} and~\eqref{eq:treatment-of-damping} in the Hamiltonian formalism~\eqref{eq:ours-explicit} gives the two mode dynamical system for skyrmion breathing
\begin{subequations}
\label{eq:reparametrised-system}
\begin{align}
\frac{dr}{d\tau}&=-\left(\frac{B}{1-|B|} \sin\eta + \alpha \frac{c_{0}c_{\eta}}{\tilde{r}_{\mathrm{eq}}}\frac{r\epsilon-1}{r^{3}} \right)  \label{eq:rdot}\\
\frac{d\eta}{d\tau}&=\left( 2\frac{r\epsilon-1}{r^{3}}-\alpha \frac{c_{0}c_{r}\tilde{r}_{\mathrm{eq}}}{2}\frac{B}{1-|B|} \sin\eta\right), \label{eq:edot-new}
\end{align}
\end{subequations}
with $\epsilon=\epsilon(r,\eta)$ as defined in Eq.~\eqref{eq:rescaled-energy} and $\tau$ is the rescaled time, 
\begin{equation}
\tau = p_{\alpha}t = \left(\frac{2 c_{0}c_{1}}{4+(\alpha c_{0})^{2}c_{\eta}c_{r}}\frac{1-|B|}{\tilde{r}_{\mathrm{eq}}}\right) t. 
\label{eq:timeRescaling}
\end{equation}

Numerical solutions of Eqs.~\eqref{eq:reparametrised-system} with and without damping are shown in Fig.~(\ref{fig:high-energy-oscillation}a) where the initial energy is set above the saddle energy. 
These results show the characteristic transition from rotating to oscillating phase in the $r$, $\eta$ and $\epsilon$ damped evolutions as compared to their undamped, constant energy analogs. 
All three variables relax towards the equilibrium state as expected. 
The rate of energy loss is qualitatively different in the two phases, transitioning from a linear-like to an exponential decay as further discussed below.
Micromagnetic simulations confirm the qualitative predictions of the effective energy model as it pertains to the skyrmion breathing modes. 
In particular, we observe the rotating (Fig.~\ref{fig:high-energy-oscillation}b) and oscillating (Fig.~\ref{fig:high-energy-oscillation}c)
breathing regimes and their distinct energy loss behavior.

\subsection{Results}

In the following we present analytical results pertaining to the expected dynamical periods, breathing amplitudes and energy decay rates. 

\subsubsection{Dynamical Periods}

\begin{figure}[t]
    \includegraphics[width=0.8\columnwidth]{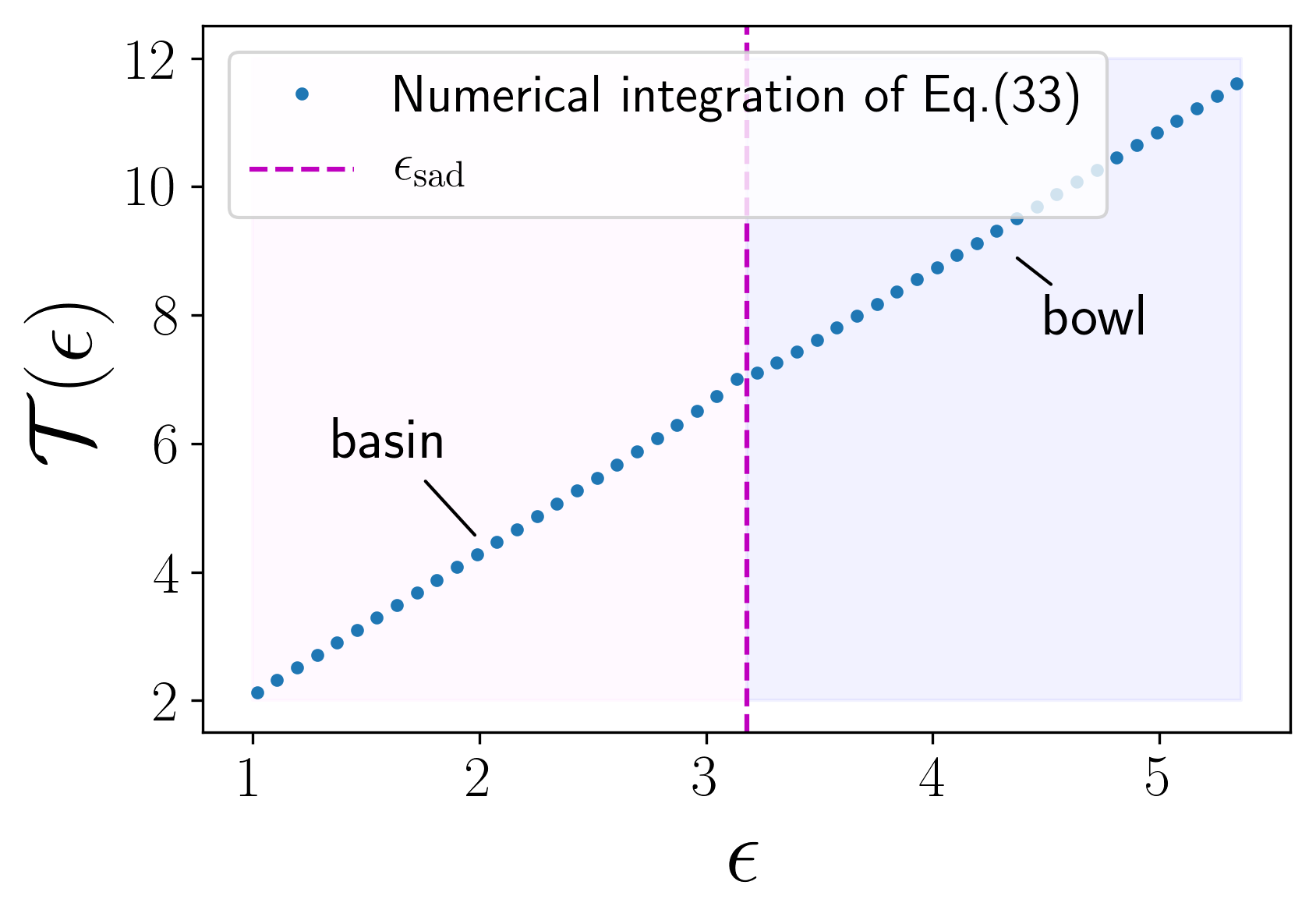}
    \caption{
    Numerical calculations of the periods $\mathcal{T}(\epsilon)$ with $B=0.82$. 
    In the regime $\epsilon>\epsilon_{\mathrm{sad}}$, the rotational `bowl' breathing modes are shown. 
    As expected from equation~\eqref{eq:Periods}, the period scales linearly with the system's energy $\mathcal{T}(\epsilon)\propto\epsilon$ in both regimes.
       }
    \label{fig:period-figure}
\end{figure}

The undamped periods of motion can be calculated from Eqs.~\eqref{eq:rdot}--~\eqref{eq:edot-new} as
\begin{widetext}
\begin{equation}
\label{eq:Periods}
\mathcal{T}(\epsilon)=\left\{\begin{array}{ll}
\mathrm{sgn}(B)(1-\left| B\right| )^2 \displaystyle\int_{\eta_{\mathrm{min}}}^{\eta_{\mathrm{max}}} d\eta \, \displaystyle\frac{2 \epsilon ^2 (1-\left| B\right| )+B \cos (\eta )-1}{(B \cos (\eta )-1)^2 \sqrt{(1-\left| B\right| ) \left(\epsilon ^2 (1-\left| B\right| )+B \cos (\eta )-1\right)}}, & \text{with} \,\epsilon<\epsilon_{\mathrm{sad}}, \\
(1-|B|)^{2}\left[ \displaystyle\int_{0}^{2\pi} d\eta \,\displaystyle\frac{2 \epsilon ^2 (1-\left| B\right| )+B \cos \eta -1}{2 (B \cos \eta -1)^2 \sqrt{(1-\left| B\right| ) \left(\epsilon ^2 (1-\left| B\right| )+B \cos \eta -1\right)}}\mp\displaystyle\frac{2\pi \epsilon}{(1-B^{2})^{3/2}} \right], & \text{with} \,\epsilon>\epsilon_{\mathrm{sad}}.
\end{array}\right.
\end{equation}
\end{widetext}
where the limits of integration are $\eta_{\mathrm{max}}=\cos^{-1}\{[1-\epsilon^{2}(1-|B|)]B^{-1}\}$ and $\eta_{\mathrm{min}} = -\eta_{\mathrm{max}} +\theta(-B)2\pi$ with $\theta(x)=1$ if $x>0$ (or $0$ otherwise) being the Heaviside function.
For the case of $\epsilon>\epsilon_{\mathrm{sad}}$ the positive and negative signs refer to bowl-like and horn-like rotations respectively. 
Direct numerical integrations of these formulas show that the period scales linearly in energy for both the rotating and oscillating breathing modes (see Fig.~\ref{fig:period-figure}). 
Our theory allows to explore dynamics beyond the small amplitude limit in an ansatz independent manner. It includes, however the small radius perturbation calculated previously in the literature derived by using a non- energy-minimizing ansatz.
For small oscillations around equilibrium the period can be computed up to $\mathcal{O}(\epsilon-1)$ as
\begin{equation}
\mathcal{T} = \pi\sqrt{2\left(\frac{1-|B|}{|B|}\right)}. 
\label{eq:small-oscillation-period}
\end{equation}
Upon converting back to physical time using~\eqref{eq:timeRescaling} and recalling from~\eqref{eq:ReqDefinition} how the physical equilibrium radius scales with the material parameters $\tilde{r}^{-1}_{\mathrm{eq}}\sim\sqrt{1-|B|}$, one recovers that the period of small oscillations around equilibrium scales as $T\propto \tilde{r}^2_{\mathrm{eq}}$ in agreement with previous literature.~\cite{Kravchuk2017a, Rodrigues2017a}

In Figure~\ref{fig:micromagnetic-matching} we emphasize the predictive power of the effective model by comparing theory to micromagnetic modeling in the undamped dynamical limit. 
Radial and angular trajectories obtained by integrating~\eqref{eq:reparametrised-system} match very well with equivalent micromagnetic simulations in both rotating (a) and oscillating (b) regimes. 
While the oscillating breathing modes match almost perfectly, a small deviation is seen between physical and predicted radial dynamics in the rotating regime.  
This is due to a breakdown in the large skyrmion approximation underpinning the theory whenever the skyrmion contracts to sizes comparable to the profile's wall width. 
To illustrate this, the inset of subfigure (a) shows the different profiles observed for maximum and minimum skyrmion radii throughout one rotational period.

\subsubsection{Breathing Amplitudes}  

The maximum and minimum possible skyrmion radii derived from the model are,\footnote{From here on we select $B>0$ for definiteness.}
\begin{subequations}
\label{eq:max-min-radiuses}
\begin{align}
r^{\pm}_{1}&=\epsilon \pm \sqrt{\epsilon^{2}-1}, \label{eq:basin-radiuses} \\
r^{\pm}_{2}&= \frac{1-B}{1+B}\epsilon\left(1\pm\sqrt{1-\frac{1+B}{(1-B)\epsilon^2}}\right),
\end{align}
\end{subequations}
which may be obtained directly from Eq.~\eqref{eq:const-E-orbit} upon setting $\eta=0,\pi$. 
At energies below the saddle energy ($\epsilon<\epsilon_{\mathrm{sad}}$), $r_{1}^{+}$ is the maximum radius of oscillations and $r_{1}^{-}$ is the minimum radius of oscillations; meanwhile, for energies above the saddle energy ($\epsilon>\epsilon_{\mathrm{sad}}$), $r_{1}^{+}$ is the maximum radius of rotations on the bowl, $r_{2}^{+}$ is the minimum radius of rotations on the bowl, $r_{2}^{-}$ is the maximum radius of rotations on the horn, and $r_{1}^{-}$ is the minimum radius of rotations on the horn. 
We see from Eq.~\eqref{eq:basin-radiuses} that the stationary points of the $r$-oscillations for breathing modes below the saddle energy are independent of the material properties. 
This may be likened to a mass-on-a-spring system where the amplitude is fully determined by the initial extension from equilibrium even though the specific dynamics connecting the two extrema of motion do depend on the size of the spring constant and the mass. 
In this case this is not a trivial consequence of small harmonic oscillations around equilibrium however because it is true for the entire $1<\epsilon<\epsilon_{\mathrm{sad}}$ range.

\subsubsection{Energy Decay Rates} 

From the expression for the skyrmion's energy Eq.~\eqref{eq:rescaled-energy} and the breathing equations of motion Eq.~\eqref{eq:reparametrised-system}, one can quantify the energy dissipated by the system as:
\begin{align}
\label{eq:Edissipation}
\frac{d\epsilon}{d\tau}  &= \frac{\epsilon r-1}{r^2}\frac{dr}{d\tau}+\frac{B}{2(1-B)}r\sin\eta\,\frac{d\eta}{d\tau} \\
		&= -\alpha\frac{c_0c_{\eta}}{\tilde{r}_{\mathrm{eq}}}\left[\frac{(\epsilon r -1)^2}{r^5}+\frac{c_r \tilde{r}_{\mathrm{eq}}^2}{c_{\eta}}\left(\frac{B}{2(1-B)}\right)^2r\sin^2\eta\right], \notag
\end{align} 
which we will use to analytically explain the distinction between linear and exponential decay observed in the numerical calculations (see Fig.~\ref{fig:high-energy-oscillation}). 
Since Eq.~\eqref{eq:Edissipation} is globally negative, except at the energy minimum $(r,\eta)=(1,0)$ where it vanishes, it correctly describes a dissipative process that relaxes the skyrmion to its equilibrium state.

\begin{figure}[t]
\includegraphics[width=\columnwidth]{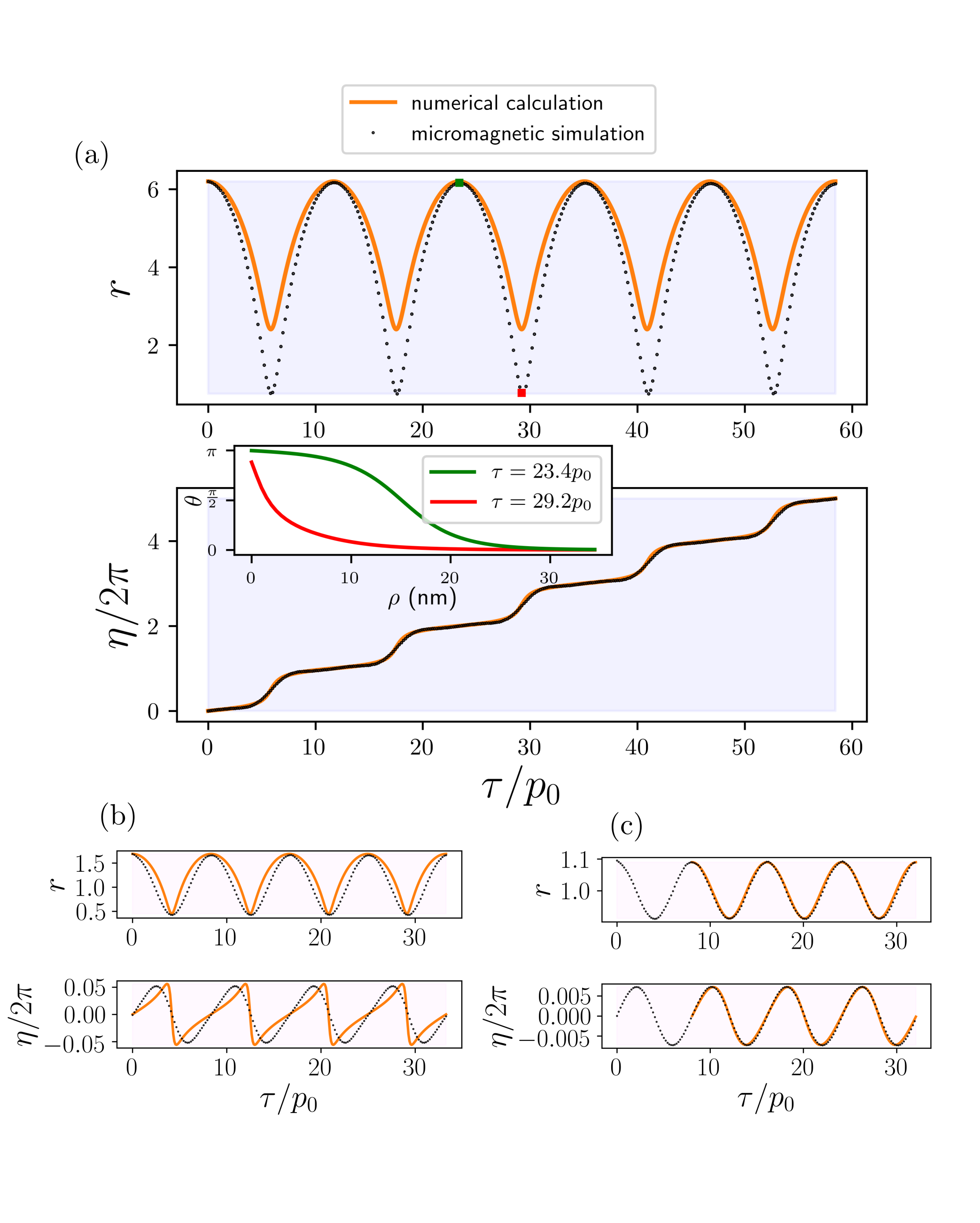}
\caption{
Comparison of micromagnetic simulations (black points) and effective model predictions (orange lines) in the undamped limit for (a) rotational breathing (light blue background) and (b)--(c) oscillating breathing (light pink background).
We find strong agreement for the weak oscillatory breathing, including the expected simple harmonic motion for small oscillations in (c) about equilibrium,  but our model shows a  deviation from the micromagnetic simulations in the rotational regime. 
The mismatch arises from the breakdown of the large skyrmion radius approximation (see inset in (a)) when the skyrmion contracts to its smallest size. 
The parameters in all three simulations are $\alpha=10^{-8}$, $D=3.2\times 10^{-3}$ Jm$^{-2}$, $A=1.5\times 10^{-11}$ Jm$^{-1}$ and $K=1.1\times 10^{6}$ Jm$^{-3}$, i.e.\ $g\approx 0.62$. 
}
    \label{fig:micromagnetic-matching}
\end{figure}

If the skyrmion's rotational and oscillatory modes precess on timescales sufficiently small compared to those for energy dissipation, the two dynamics can be effectively decoupled by averaging~\eqref{eq:Edissipation} over constant energy orbits to obtain a single ordinary differential equation describing the energy lost by the system over time. 
One then has:
\begin{align}
\label{eq:energydecay}
\left\langle\frac{d\epsilon}{d\tau}\right\rangle=-\frac{\alpha}{\mathcal{T}(\epsilon)}\frac{c_0c_{\eta}}{\tilde{r}_{\mathrm{eq}}}&\left[\oint\mathrm{d}\eta\,\frac{(\epsilon r(\eta) -1)^2}{r^5(\eta)}\right. \\
&\left. + \frac{c_r \tilde{r}_{\mathrm{eq}}^2}{c_{\eta}}\left(\frac{B}{2(1-B)}\right)^2\oint\mathrm{d}\eta\,r(\eta)\sin^2\eta\right], \notag
\end{align} 
where the integrals are performed over one complete oscillatory/rotational orbits~\eqref{eq:const-E-orbit}. 
In what follows we will leverage the fact that orbital periods~\eqref{eq:Periods} scale linearly with the orbit's energy $T(\epsilon)\propto \epsilon$ (see Fig.~\ref{fig:period-figure}). 
The above integrals are not exactly solvable but a series of upper and lower bounds can still be constructed (see Appendix \ref{app:ExtraEnergyAverages}). 
For the rotational breathing mode, one can show that:
\begin{widetext}
\begin{align}
\label{eq:ErotBound}
\left\langle\frac{d\epsilon}{d\tau}\right\rangle\vert_{\mathrm{rot}}&\ge-\frac{\alpha}{\mathcal{T}(\epsilon)}\frac{c_0c_{\eta}}{\tilde{r}_{\mathrm{eq}}}\left[\frac{2\pi\epsilon_{\mathrm{sad}}^6}{\epsilon}+2\pi\epsilon\frac{c_r \tilde{r}_{\mathrm{eq}}^2}{4c_{\eta}}\left(\frac{B}{1-B}\right)^2\right]\simeq -\alpha\left[C_1+\frac{C_2}{\epsilon^2}\right],
\end{align}
\end{widetext}
where we absorb all constants into $C_{1,2}>0$. 
By construction, the solution of this new bounding equation is guaranteed to decay faster than the true solution of Eq.~\eqref{eq:Edissipation}. 
Since $\epsilon\gg 1$ for rotational modes, this upper bound guarantees at most a linear decay to the skyrmion's energy.

Following a similar reasoning for the oscillatory breathing modes by constructing a lower bound to the energy dissipation rate, one finds (see Appendix \ref{app:ExtraEnergyAverages}),
\begin{widetext}
\begin{equation}
\label{eq:EoscBound}
\left\langle\frac{d\epsilon}{d\tau}\right\rangle\vert_{\mathrm{osc}}\le-\frac{\alpha}{\mathcal{T}(\epsilon)}\frac{c_0c_{\eta}}{\tilde{r}_{\mathrm{eq}}}\left[\frac{2}{\epsilon^2}\frac{c_r \tilde{r}_{\mathrm{eq}}^2}{4c_{\eta}}\left(\frac{B}{1-B}\right)^{5/2}\left(\sin\eta_+(\epsilon)\cos^2\eta_+(\epsilon)-\cos\eta_+(\epsilon)\right)\right]
\simeq -\alpha C_3\,\epsilon\sqrt{\epsilon-1},
\end{equation}
\end{widetext}
where $\eta_+(\epsilon)=\cos^{-1}\{[1-\epsilon^{2}(1-B)]B^{-1}\}$ is the maximum range of domain wall tilt angle attained during a single constant energy oscillation. 
Since the solution of~\eqref{eq:EoscBound} is guaranteed to decay slower than that of (\ref{eq:Edissipation}), the true energy loss in the oscillating regime must decay at least exponentially. 
These arguments confirm the sharp transition observed in the dissipation rate when the skyrmion breathing dynamics cross the saddle separatrix when transitioning between $\epsilon>\epsilon_{\mathrm{sad}}$ and $\epsilon<\epsilon_{\mathrm{sad}}$ states, see Fig.~\eqref{fig:high-energy-oscillation}.

\section{Conclusions}

In this work we have derived the Hamiltonian system for the low energy excitations of rotationally symmetric magnetic (anti-)skyrmions in an ansatz-independent manner. 
By means of scaling and symmetry arguments we modeled the breathing mode of (anti-)skyrmions in terms of collective coordinates, where the area of reversed spins in the skyrmion core and the skyrmion phase are conjugated variables in phase space. 
As seen from the form of the energy landscape, our model exhibits a rich behavior which is confirmed by micromagnetic simulations of (anti-)skyrmion structures in magnetic substrates with translational invariance along the anisotropy axis. 
The main results presented in this paper include the analytical and numerical investigation of the well-known oscillatory breathing mode where small amplitude oscillations in the radius and skyrmion phase around equilibrium proceed in tandem, as well as the description of rotational breathing behavior, characterized by large radius oscillations and a continuous non-uniform precession of the phase. 
Furthermore, we predict two distinct regimes of energy dissipation where the average power loss of large amplitude rotational breathing modes decays linearly as opposed to exponentially for the oscillating modes. 
We expect that these distinctive energy decays will allow to detect the different modes experimentally. 
It must be stated that the limit of our model lies in the implicit assumption of fixed skyrmion wall profiles. 
The next order approximation would be to incorporate the skew of the wall profile by introducing an additional pair of collective coordinates. 
Doing so would allow extension of this analysis to skyrmion radii much smaller than those allowed by this work.
We would like to emphasize that the results described here hold for both skyrmions and antiskyrmions. 
Therefore a perfect test system will be one where both of them occur simultaneously. 
This is for example naturally the case when skyrmion and antiskyrmion pairs are created,\cite{Everschor-Sitte2016, Leonov2016c, Stier2017} or in systems with certain symmetries.~\cite{Hoffmann2017}

\section{Acknowledgments}

The groups at Mainz acknowledge funding from the Transregional Collaborative Research Center (SFB/TRR) 173 SPIN+X, the Graduate School of Excellence 
Materials Science in Mainz (MAINZ, GSC 266), the German Research Foundation (DFG) under the Project No. EV 196/2-1 and the Alexander von Humboldt Foundation.

B.~M.~, D.~R.~ and D.~P.~ contributed equally to this work.

\appendix

\section{Poisson brackets}
\label{app:PBs}
In the main text we applied techniques from Hamiltonian mechanics. Poisson brackets entered at the level of the collective coordinates where we were concerned with just time dependent functions, and also at the level of the LLG \textit{field} equation where the magnetization field depends on both space and time. Below we review equations for Poisson brackets relevant to this work.

\subsection*{Time dependent functions}
The Poisson bracket convention for time dependent functions $A(t)$ and $B(t)$ is 
 \begin{equation}
\{A,B\}_{\vect{q},\vect{p}}=\sum_{i}\left(\frac{\partial A}{\partial q_{i}}\frac{\partial B}{\partial p_{i}} -\frac{\partial A}{\partial p_{i}}\frac{\partial B}{\partial q_{i}}\right).
\label{eq:PBs}
\end{equation}
corresponding to the simple action $\mathcal{S}=\int dt (\sum_{i}p_{i}\dot{q}_{i}-H)$. For example, a basic result is the Poisson bracket between the canonical coordinates and momenta, 
\begin{equation}
\{q_{i},p_{j}\}_{\vect{q},\vect{p}}=\delta_{ij}.
\end{equation}
Using the Poisson bracket, the time derivative of any function $f(\vect{q},\vect{p},t)$ can be calculated,

\begin{align}
\dot{f}&= \sum_{i}\left(\frac{\partial f}{\partial q_{i}}\dot{q}_{i}+\frac{\partial f}{\partial p_{i}}\dot{p}_{i}\right)+ \frac{\partial f}{\partial t} \notag \\
&=\{f,H \}_{\vect{q},\vect{p}}+\frac{\partial f}{\partial t},
\label{eq:PB-t-derivative-simple}
\end{align}
where Hamilton's equations $\dot{p_{i}}=-\frac{\partial H}{\partial q_{i}}, \quad \dot{q_{i}}=\frac{\partial H}{\partial p_{i}}$ were used in the second line.

A further rule can be derived for the Poisson bracket between quantities that have a known dependence on functions of the canonical variables, e.g.\ $A(t)=A[f_{1}(\vect{q},\vect{p},t), f_{2}(\vect{q},\vect{p},t),\dots]$ and similar for $B(t)$:
\begin{equation}
\{A,B\}_{\vect{q},\vect{p}}=\sum_{i,j}\frac{\partial A}{\partial f_{i}}\frac{\partial B}{\partial f_{j}}\{f_{i}, f_{j}\}_{\vect{q},\vect{p}}.
\label{eq:ABPb}
\end{equation}

\subsection*{Fields}
There is an analogous description to Eqs.~\eqref{eq:PBs}--~\eqref{eq:ABPb} for fields. The Poisson bracket between scalar fields $\mathcal{A}(\vect{x})$ and $\mathcal{B}(\vect{x}')$, is
\begin{align}
&\{\mathcal{A}(\vect{x}),\mathcal{B}(\vect{x}')\}_{\vect{\Phi},\vect{\Pi}}= \notag \\
&\sum_{i} \int \,d \vect{y}\left(\frac{\delta \mathcal{A}(\vect{x})}{\delta \Phi_{i}(\vect{y})}\frac{\delta \mathcal{B}(\vect{x}')}{\delta \Pi_{i}(\vect{y})}-\frac{\delta \mathcal{A}(\vect{x})}{\delta \Pi_{i}(\vect{y})}\frac{\delta \mathcal{B}(\vect{x}')}{\delta \Phi_{i}(\vect{y})}\right)
\label{eq:PB-def-for-fields}
\end{align}
corresponding to an action of the form $\mathcal{S}=\int dt \int dV (\sum_{i}\Pi_{i}\partial_{t}\Phi_{i}-\mathcal{H})$. 
So the Poisson bracket between the canonical fields is
\begin{equation}
\{\Phi_{i}(\vect{x}),\Pi_{j}(\vect{x}')\}_{\vect{\Phi},\vect{\Pi}}=\delta_{ij}\delta(\vect{x}-\vect{x}').
\end{equation}
The time derivative for a field $\mathcal{F}(\vect{x},t)$, similar to Eq.~\eqref{eq:PB-t-derivative-simple}, is 
\begin{equation}
\dot{\mathcal{F}}=\{\mathcal{F},H \}_{\vect{\Phi},\vect{\Pi}}+\frac{\partial \mathcal{F}}{\partial t}
\label{eq:field-time-derivative}
\end{equation}
by use of Hamilton's equations $\dot{\Phi}_{i}=\frac{\delta H}{\delta \Pi_{i}}$ and $\dot{\Pi}_{i}=-\frac{\delta H}{\delta \Phi_{i}}$. 
Finally the useful identity in analogy to Eq.~\eqref{eq:ABPb}, for the Poisson bracket between quantities that have known dependence on a set of functions, say $\{g_{i}(\vect{x})\}$, is
\begin{align}
&\{\mathcal{A}(\vect{x}),\mathcal{B}(\vect{x}')\}_{\vect{\Phi},\vect{\Pi}}= \notag \\
&\sum_{i,j} \int \,d \vect{y}\int\,d\vect{z}\frac{\delta \mathcal{A}(\vect{x})}{\delta g_{i}(\vect{y})}\frac{\delta \mathcal{B}(\vect{x}')}{\delta g_{j}(\vect{z})}
\{g_{i}(\vect{y}),g_{j}(\vect{z})\}_{\vect{\Phi},\vect{\Pi}}.
\label{eq:PB-for-fields}
\end{align}

\subsection*{Example: Hamiltonian formulation of the LLG}

The LLG Eq.~\eqref{eq:LLG} may be written as a Hamiltonian Eq.~\eqref{eq:LLG-Hamiltonian-form} if one assumes that the local magnetization obeys the algebra
\begin{equation}
\{m_{i}(\vect{x}),m_{j}(\vect{x}')\}_{\Phi,\Pi}=-\frac{1}{\mathcal{J}}\sum_{k}\epsilon_{ijk}m_{k}(\vect{x})\delta(\vect{x}-\vect{x}').
\label{eq:appendix-m-component-PBs}
\end{equation}
where $\mathcal{J}=M_{s}/\gamma_{0}$. This may be verified explicitly. For example, using the spherical parameterization of the magnetization employed in the main text, we may identify the two canonical fields as $\Phi(\vect{x})=\phi(\vect{x})$ and $\Pi(\vect{x})=\mathcal{J}(1-\cos\theta(\vect{x}))$ from the spin Berry phase action Eq.~\eqref{eq:Spin-BP-action}. Then the three non-zero Poisson brackets between the magnetization components $m_{x}$, $m_{y}$ and $m_{z}$ in Eq.~\eqref{eq:appendix-m-component-PBs} are straightforwardly verified using Eq.~\eqref{eq:PB-def-for-fields} with these canonical fields. Moreover in this situation Eq.~\eqref{eq:PB-for-fields} reduces to a cross product structure
\begin{align}
\{\mathcal{A}(\vect{x}),\mathcal{B}(\vect{x}')\}_{\Phi,\Pi}
=-\frac{1}{\mathcal{J}}\int d \vect{y}\,\vect{m}(\vect{y})\cdot\left(\frac{\delta \mathcal{A}(\vect{x})}{\delta \vect{m}(\vect{y})} \times \frac{\delta \mathcal{B}(\vect{x}')}{\delta \vect{m}(\vect{y})}\right).
\end{align}
Hence, by using Eqs.~\eqref{eq:field-time-derivative}--\eqref{eq:appendix-m-component-PBs}, evaluating the time evolution of a component $m_{l}(\vect{x})$ gives,
\begin{align}
&\{m_{l}(\vect{x}),H \}_{\Phi,\Pi} \notag \\
&= \sum_{i,j} \int \,d\vect{y}\int\,d\vect{z} \frac{\delta m_{l}(\vect{x})}{\delta m_{i}(\vect{z})}\frac{\delta H}{\delta m_{j}(\vect{y})}\{m_{i}(\vect{z}),m_{j}(\vect{y}) \}_{\Phi,\Pi} \notag \\
&= -\frac{1}{\mathcal{J}}\sum_{jk}\epsilon_{ljk}\frac{\delta H}{\delta m_{j}(\vect{x})}m_{k}(\vect{x}) \notag \\
&= \frac{1}{\mathcal{J}}\left[\vect{m}(\vect{x})\times \frac{\delta H}{\delta \vect{m}(\vect{x}) }\right]_{l},
\end{align}
which is the anticipated precessional term for undamped motion.

\section{Comparison between generalized Thiele method and the Hamiltonian formalism}
\label{app:hamiltonian-thiele-comparison}

The generalized Thiele method is based on the idea of describing the dynamics of certain magnetic configurations just in terms of the time evolution of a finite number of collective coordinates describing the \textit{soft modes},~\cite{Tretiakov2008,Clarke2008} for which we have that,
\begin{equation}
\dot{\vect{m}} = \sum_{i=1}^{2N}\left(\frac{\partial\vect{m}}{\partial\xi_{i}}\right)\dot{\xi}_{i}.
\label{eq:time-expansion}
\end{equation}
This decomposition has been successfully applied in the description of the low energy excitations of topological magnetic textures. For example, for the field-driven or current-driven motion of domain walls the physics is well described up to a certain magnitude of the applied driving field or current~\cite{Tatara2004} by a soft mode described by the two collective coordinates: the domain wall position and the tilt angle of the magnetization inside the wall. Another example is the dynamics of  the position $(X,Y)^{\intercal}$ of a rigid homogenous domain in steady translational motion as described by Thiele.~\cite{Thiele1973}
Considering the expansion Eq.~\eqref{eq:time-expansion}, performing the projection of the LLG  Eq.~\eqref{eq:LLG} onto $\vect{m}\times \partial_{\xi_{i}}\vect{m}$ and integrating over volume gives a generalization of Thiele's equations~\cite{Thiele1973,Tretiakov2008,Everschor2011}
\begin{equation}
\sum_{j} \mathcal{G}_{ij}\dot{\xi}_{j}=\frac{1}{\mathcal{J}}\frac{\partial H}{\partial \xi_{i}}+\alpha \sum_{j} \mathcal{D}_{ij}\dot{\xi}_{j}
\label{eq:generalized-thiele}
\end{equation}
where the matrix elements $\mathcal{G}_{ij}$ and $\mathcal{D}_{ij}$ were defined in Eqs.~\eqref{eq:Giro-tensor} and \eqref{eq:Visco-tensor} respectively.
By assuming that the matrix $\mathcal{G}_{ij}[\vect{\xi}]$ is invertible, one may also write
\begin{equation}
\dot{\xi}_{k}=\sum_{i,j}\mathcal{G}^{-1}_{ki}\mathcal{G}_{ij}\dot{\xi}_{j}=\frac{1}{\mathcal{J}}\sum_{i}\mathcal{G}^{-1}_{ki}\frac{\partial H}{\partial \xi_{i}} +\alpha\sum_{i,j}\mathcal{G}^{-1}_{ki} \mathcal{D}_{ij}\dot{\xi}_{j},
\end{equation}
where $\mathcal{G}_{ki}^{-1}\equiv (\mathcal{G}^{-1})_{ki}$ are the elements of the inverse matrix.
Comparing this equation with the Hamiltonian Eq.~\eqref{eq:xi-dot}, we obtain that the equivalence between the generalized Thiele approach and Hamiltonian approach is embedded in the following identities
\begin{subequations}
\begin{equation}
\{\xi_{i},\xi_{j} \}_{\vect{\xi},\vect{p}_{\vect{\xi}}}\equiv \mathcal{J}^{-1}\mathcal{G}^{-1}_{ij},
\label{eq:Poisson-Bracket-Thiele}
\end{equation}
\begin{equation}
\gamma_{\xi_{k}} \equiv \alpha \sum_{i,j}\mathcal{G}^{-1}_{ki}\mathcal{D}_{ij}\dot{\xi}_{j}.
\end{equation} 
\end{subequations}
The result for the dissipative term $\gamma_{\xi_{j}}$ in terms of the Poisson brackets $\{\xi_{i},\xi_{j}\}_{\vect{\xi},\vect{p}_{\vect{\xi}}}$ is also derived in a more general manner (see the main text) and therefore this structure is general for including viscous damping into Hamilton's equations, regardless of the system of study.

To further illustrate the Hamiltonian approach, we present as an example the Poisson bracket for $X$ and $Y$, describing the position of a skyrmion, for the soft mode associated with translational motion.

\subsubsection*{Example: Derivation of the $X,Y$ Poisson bracket for the translational modes from the spin Berry phase action}
\label{app:XY-PB-derivation}

By virtue of Thiele's traveling wave ansatz, $\vect{m}(\vect{r},t)=\vect{m}_{0}[\vect{r}-\vect{r}_{s}(t)]$, the unit magnetization has the properties $\dot{\vect{m}}=-(\dot{\vect{r}}_{s}\cdot\nabla)\vect{m}$ and $\partial_{X,Y}\vect{m}=-\partial_{x,y}\vect{m}$ where $\vect{r}_{s}=(X,Y)^\intercal$. 
The Poisson brackets between $X$ and $Y$, and hence access to the conservative dynamics, are derived by expanding the spin Berry phase action $S_{B}=\int dt L_{B}$ around $\vect{r}_{s}=0$, while making use of these properties. The Lagrange function here is
\begin{align}
L_{B}&=\int dV \mathcal{J}(1-\cos\theta)\dot{\phi} \notag \\
&= \int dV \mathcal{J}(1-\cos\theta)(\dot{X}\partial_{X}\phi + \dot{Y}\partial_{Y}\phi)  
\end{align}
Performing the expansion, denoting $\theta=\theta[\vect{r},\vect{r}_{s}]$ and $\phi=\phi[\vect{r},\vect{r}_{s}]$, this becomes to lowest order in $X, Y, \dot{X}$ and $\dot{Y}$
\begin{widetext}
\begin{align}
L_{B}&\approx \int dV \mathcal{J}\left[(1-\cos\theta[\vect{r},0])+X\partial_{x}\cos\theta[\vect{r},0]+Y\partial_{y}\cos\theta[\vect{r},0]\right]\left[\dot{X}(-\partial_{x}\phi[\vect{r},0] )+ \dot{Y}(-\partial_{y}\phi[\vect{r},0]) \right] \notag \\
&=\mathcal{J}\left\{ -\dot{X} \left[\int dV(1-\cos\theta[\vect{r},0])\partial_{x}\phi[\vect{r},0] \right] -\dot{Y} \left[\int dV(1-\cos\theta[\vect{r},0])\partial_{y}\phi[\vect{r},0] \right] \right . \notag \\
& -X\dot{X}\left[\int dV \partial_{x}\cos\theta[\vect{r},0]\partial_{x}\phi[\vect{r},0] \right]-Y\dot{Y}\left[\int dV \partial_{y}\cos\theta[\vect{r},0]\partial_{y}\phi[\vect{r},0] \right]  \notag \\
& \left.-X\dot{Y}\left(\int dV \partial_{x}\cos\theta[\vect{r},0]\partial_{y}\phi[\vect{r},0] \right)-Y\dot{X}\left(\int dV \partial_{y}\cos\theta[\vect{r},0]\partial_{x}\phi[\vect{r},0] \right)  \right\}.
 \end{align}
\end{widetext}
The first four terms in the expansion can be written as total derivatives of $X$ or $Y$ or their squares and therefore do not enter into the equations of motion. 
Next, by integrating the final term by parts in the action with respect to time the boundary contribution similarly vanishes and it follows that the spin Berry phase action reduces to lowest order to
\begin{align}
S_{B}&= \int dt\mathcal{J}\left\{X\dot{Y}\int dV \left(\partial_{y}\cos\theta[\vect{r},0]\partial_{x}\phi[\vect{r},0]  \right. \right. \notag \\
&\left. \left.- \partial_{x}\cos\theta[\vect{r},0]\partial_{y}\phi[\vect{r},0] \right)\right\}.
\label{eq:S-Berry-XY-step}
\end{align}
This is Eq.~\eqref{eq:SbXY} of the main text. Finally, by integrating over $z$ from $0$ to $z_{0}$ and comparing Eq.~\eqref{eq:S-Berry-XY-step} to the topological charge,
\begin{align}
\mathcal{Q} &= \frac{1}{4\pi}\int dx dy\,  \vect{m}\cdot(\partial_{x}\vect{m}\times\partial_{y}\vect{m}) \notag \\
&= \frac{1}{4\pi} \int dx dy \, \sin\theta (\partial_{y}\phi \partial_{x}\theta - \partial_{y}\theta \partial_{x}\phi) \notag \\
&= \frac{1}{4\pi} \int dx dy\,(\partial_{y}\cos\theta \partial_{x}\phi - \partial_{y}\phi \partial_{x}\cos\theta) \label{eq:Q-polar-coords},
\end{align}
one finds the effective action $S_\text{eff}=\int dt (4\pi Q z_{0}\mathcal{J}X\dot{Y}-H)$. This gives the Poisson bracket $\{Y,X\}_{Y,p_{Y}}=(4\pi Q z_{0}\mathcal{J})^{-1}$, in the main text (see Eq.~\eqref{eq:XY-PB}).

\section{Generalizations of the effective model}
\label{app:ExtraSecIV}
The arguments used to expand the micromagnetic model Eq.~\eqref{eq:efunctional} in powers of $R$ can be used for other interactions, including dipole-dipole interactions and bulk DMI. In this note we briefly outline how to modify the effective energy Eq.~\eqref{eq:effectiveE} to take into account these terms.

The dipole-dipole interaction is known to produce in thin-films with strong perpendicular magnetic anisotropy to modify the strength of the anisotropy,~\cite{Hubert2009} $K_{\mathrm{eff}} = K -\frac{1}{2}\mu_{0}M_{s}^2$. Moreover, for circular skyrmions, it also produces a coupling for the $\eta$ angle. From symmetry arguments, we argue that this interaction is invariant under the transformation $\eta \rightarrow -\eta$. And, from the scaling argument, we argue that it decays with the inverse of the radius. Therefore, the contribution to the $\eta$ coupling may be written as $-c_{dd}(\cos^2 \eta)/r$ where $c_{dd}$ depends on the exact profile of the skyrmion. Usually $c_{dd}$ is at least an order of magnitude smaller than the other $c$'s in the effective energy Eq.~\eqref{eq:effectiveE}. This contribution was calculated using the domain wall ansatz in a previous paper.~\cite{Knoester2014}

Another example of possible modifications to the energy is the generalization of the DMI to include bulk and hybrid DMI.  In the case of hybrid DMI we take into account a combination of bulk and interfacial DMI.~\cite{Kim2018} The general contribution to the energy density becomes
\begin{align}
\mathcal{\tilde{H}}_{\mathrm{DMI}} &= D^{\mathcal{N}}(m_{z}\partial_{x}m_{x}-m_{x}\partial_{x}m_{z}) \notag \\
&\pm D^{\mathcal{N}}(m_{z}\partial_{y}m_{y}-m_{y}\partial_{y}m_{z}) \notag \\
&-D^{\mathcal{B}}(m_{z}\partial_{y}m_{x}-m_{x}\partial_{y}m_{z}) \notag \\
&\pm D^{\mathcal{B}}(m_{z}\partial_{x}m_{y}-m_{y}\partial_{x}m_{z}),
\end{align}
where $\mathcal{B}$ and $\mathcal{N}$ stand for bulk and interfacial DMI respectively, and the different signs correspond to the energies that stabilizes skyrmions and antiskyrmions, as stated in the main text. In this case, the effective contribution is given by
\begin{equation}
\tilde{E}_{\mathrm{DMI}}(\tilde{r},\eta) = -\tilde{r}(g^{\mathcal{N}}\cos\eta + g^{\mathcal{B}}\sin\eta).
\end{equation}
where $g^{\mathcal{B},\mathcal{N}} = \pi D^{\mathcal{B},\mathcal{N}}/4\sqrt{AK}$. If neither $g^{\mathcal{B}}$ or $g^{\mathcal{N}}$ is zero, it produces an equilibrium $\eta$ that is different from the usual N\'eel and Bloch skyrmion and can take any value between $0$ and $2\pi$. The new equilibrium angle is given by the ratio of $g^{\mathcal{N}}$ and $g^{\mathcal{B}}$, i.e.\ $\eta = \arctan(g^{\mathcal{B}}/g^{\mathcal{N}})$.

For the case where we consider that the domain wall width $d$ is also a function of the radius, we need to keep its explicit dependence in the effective energy. In this case, by analyzing the scaling factors before the reparametrization by the domain wall width $\Delta$ it is possible to obtain the effective energy for the model Eq.~\eqref{eq:efunctional},
 \begin{equation}
 \label{eq:EffEK}
\tilde{E}(\tilde{r},\eta) = \left(\frac{c_{12}}{d} + c_{11}d - c_{3} g \cos\eta\right) \tilde{r} + \frac{c_{21}d}{\tilde{r}} ,
 \end{equation}
where $c_{12}$ and $c_{21}$ are contributions from the exchange interaction, $c_{11}$ is due to the anisotropy interaction and $c_{3}$ is due to DMI. 
The dependence of $d$ on the DMI strength may be obtained in two ways. 
One method is by analyzing the scaling behavior for a rotationally symmetric solution of the LLG equation in two dimensions.
An explicit method is to consider $r, \eta$ and $d$ as collective variables and minimizing the energy for these three parameters. Thereby we obtain,
\begin{equation}\label{eq:SkyWidth}
d = \frac{c_{3}|g|}{2c_{11}}.
\end{equation}
An important remark is that the effective description mentioned in this paper is only valid for the case that $\tilde{r} \gg d$. If the radius of the skyrmion becomes comparable to the width of the circular domain wall, it is necessary to consider the solutions for small skyrmions studied in Refs.~[\onlinecite{Bogdanov1989}, \onlinecite{Bogdanov1994}].

The effective energy Eq.~\eqref{eq:effectiveE} gives the static properties obtained in Ref.~[\onlinecite{Rohart2013}] with $c$ values independent of DMI, while the model obtained in Eq.~\eqref{eq:EffEK} with the width of the skyrmion given by Eq.~\eqref{eq:SkyWidth} corresponds to the one obtained in Ref.~[\onlinecite{Kravchuk2017a}]. For the circular domain wall ansatz, the values of the constants are given by: $c_{3} = 2$, $c_{1}=2$, and all other constants $c$ are equal to $1$. The conversion between the constants in the two approaches (Eq.~\eqref{eq:effectiveE} and Eq.~\eqref{eq:EffEK}) is given by,
\begin{align}
c_{1} &= \frac{2c_{11}c_{12}}{c_{3}|g|} + \frac{c_{3}}{2}|g|,\notag\\
c_{2} &= \frac{c_{21}c_{3}|g|}{2c_{11}}.
\end{align}

\section{Constant Energy Orbit Averages}
\label{app:ExtraEnergyAverages}

To estimate the average energy decay during the skyrmion breathing mode, one can construct upper and lower bounds for the two integrals entering in Eq.~\eqref{eq:energydecay}:
\begin{subequations}
\begin{align}
\label{appeq:integraldefs}
I_1(\epsilon;B)&\equiv\oint\mathrm{d}\eta\,\frac{(\epsilon r(\eta) -1)^2}{r^5(\eta)}\\
I_2(\epsilon;B)&\equiv\oint\mathrm{d}\eta\,r(\eta)\sin^2\eta.
\end{align}
\end{subequations}
These integrals need to be performed over one full rotational/precessional constant energy orbit as defined by the orbit trajectory:
\begin{align}
r_{\mathrm{rot}}(\eta;\epsilon)&=\frac{1-B}{2(1-B\cos\eta)}\left[1+\sqrt{1-\frac{1-B\cos\eta}{(1-B)\epsilon^2}}\,\right] \\
r^{\pm}_{\mathrm{osc}}(\eta;\epsilon)&=\frac{1-B}{2(1-B\cos\eta)}\left[1\pm\sqrt{1-\frac{1-B\cos\eta}{(1-B)\epsilon^2}}\,\right],
\end{align}
where, as discussed in the main text, $r_{\mathrm{rot}}(\eta)$ is defined for $\eta\in[0,2\pi]$ whereas the two branches $r^{\pm}_{\mathrm{osc}}(\eta)$ are only defined for $\eta\in[-\eta_+,\eta_+]$ with
\begin{equation}
\cos\eta_+=\frac{1-(1-B)\epsilon^2}{B}.
\end{equation}
For rotational phase integrals we have
\begin{equation}
\oint\mathrm{d}\eta\,f(\eta)=2\int_0^{\pi}\mathrm{d}\eta\,f(\eta),
\end{equation}
whereas for oscillatory phase integrals we have
\begin{align}
\oint\mathrm{d}\eta\,f^{\pm}(\eta)&=\int_{-\eta_+}^{\eta_+}\mathrm{d}\eta\,f^+(\eta)+\int_{\eta_+}^{-\eta_+}\mathrm{d}\eta\,f^-(\eta)\notag \\
&=2\int_0^{\eta_+}\mathrm{d}\eta\,\left(f^+(\eta)-f^-(\eta)\right),
\end{align}
one can therefore write~\eqref{appeq:integraldefs} explicitly as,
\begin{align}
I^{\mathrm{rot}}_1(\epsilon;B)&= \epsilon^2 J^{\mathrm{rot}}_3-2\epsilon J^{\mathrm{rot}}_4+J^{\mathrm{rot}}_5 \\
I^{\mathrm{rot}}_2(\epsilon;B)&= 2(1-B)\epsilon \notag \\
&\times \int_0^{\pi}\mathrm{d}\eta\,\frac{\sin^2\eta}{1-B\cos\eta}\left(1+\sqrt{1-\frac{1-B\cos\eta}{(1-B)\epsilon^2}}\right) 
\end{align}
for integrals in the rotational regime, and
\begin{align}
I^{\mathrm{osc}}_1(\epsilon;B)&= \epsilon^2 J^{\mathrm{osc}}_3-2\epsilon J^{\mathrm{osc}}_4+J^{\mathrm{osc}}_5 \\
I^{\mathrm{osc}}_2(\epsilon;B)&= 4\sqrt{B(1-B)}\notag \\
&\times \int_0^{\eta_+}\mathrm{d}\eta\,\frac{\sin^2\eta}{1-B\cos\eta}\sqrt{\cos\eta-\cos\eta_+},
\end{align} 
for integrals in the oscillatory regime, where we have further defined:
\begin{align}
J^{\mathrm{rot}}_k &\equiv \frac{2}{(1-B)^k\epsilon^k}\int_0^{\pi}\mathrm{d}\eta\,(1-B\cos\eta)^k\notag \\
&\times\left[1+\sqrt{1-\frac{1-B\cos\eta}{(1-B)\epsilon^2}}\right]^{-k},
\end{align}
and,
\begin{align}
J^{\mathrm{osc}}_k &\equiv \frac{4}{(1-B)^k\epsilon^k} \int_0^{\eta_+}\mathrm{d}\eta\,(1-B\cos\eta)^k \notag \\
&\times \left\{\left[1+\sqrt{1-\frac{1-B\cos\eta}{(1-B)\epsilon^2}}\right]^{-k} \right. \notag \\
&\left.-\left[1-\sqrt{1-\frac{1-B\cos\eta}{(1-B)\epsilon^2}}\right]^{-k}\right\}. 
\end{align} 

The necessary upper bounds to the $I^{\mathrm{rot}}_1$ and $I^{\mathrm{rot}}_2$ are obtained by approximating $1+\sqrt{1-\frac{1-B\cos\eta}{(1-B)\epsilon^2}}\leq 2$ and writing
\begin{eqnarray}
\int_0^{\pi}\mathrm{d}\eta\,\frac{\sin^2\eta}{1-B\cos\eta}\leq \frac{\pi}{2(1-B)} \\
\int_0^{\pi}\mathrm{d}\eta\,(1-B\cos\eta)^k\leq \pi(1+B)^k.
\end{eqnarray}
These bounds are independent of the energy of the orbit thus justifying Eq.~\eqref{eq:ErotBound} of the main text.

For the oscillatory phase integrals, one just focuses on $I^{\mathrm{osc}}_2$ as $J^{\mathrm{osc}}_k$ is always bounded from below by $0$. One finds:
\begin{widetext}
\begin{equation}
I^{\mathrm{osc}}_2(\epsilon;B) \geq \frac{4}{\epsilon^2}\sqrt{\frac{B}{1-B}}\int_0^{\eta_+}\mathrm{d}\eta\,\sin^2\eta(\cos\eta-\cos\eta_+) \geq\frac{2}{\epsilon^2}\sqrt{\frac{B}{1-B}}\left(\sin\eta_+\cos^2\eta_+-\cos\eta_+\right),
\end{equation}
\end{widetext}
where from the definition of $\eta_+(\epsilon)$, the term $\sin\eta_+\cos^2\eta_+$ can be expanded to dominant order in $\epsilon$ as,
\begin{widetext}
\begin{align}
\sin\eta_+\cos^2\eta_{+}=\sqrt{1-\left(\frac{1-(1-B)\epsilon^2}{B}\right)} \left(\frac{1-(1-B)\epsilon^2}{B}\right)^2 \sim 2\sqrt{\frac{(1-B)^5}{B}}\epsilon^{4}\sqrt{\epsilon-1}\left[1+\mathcal{O}(\epsilon-1)\right],
\end{align}
allowing for the reconstruction of result Eq.~\eqref{eq:EoscBound} in the main text.
\end{widetext}

\section{Details on micromagnetic simulations}

\begin{figure}[t]
    \includegraphics[width=0.8\columnwidth]{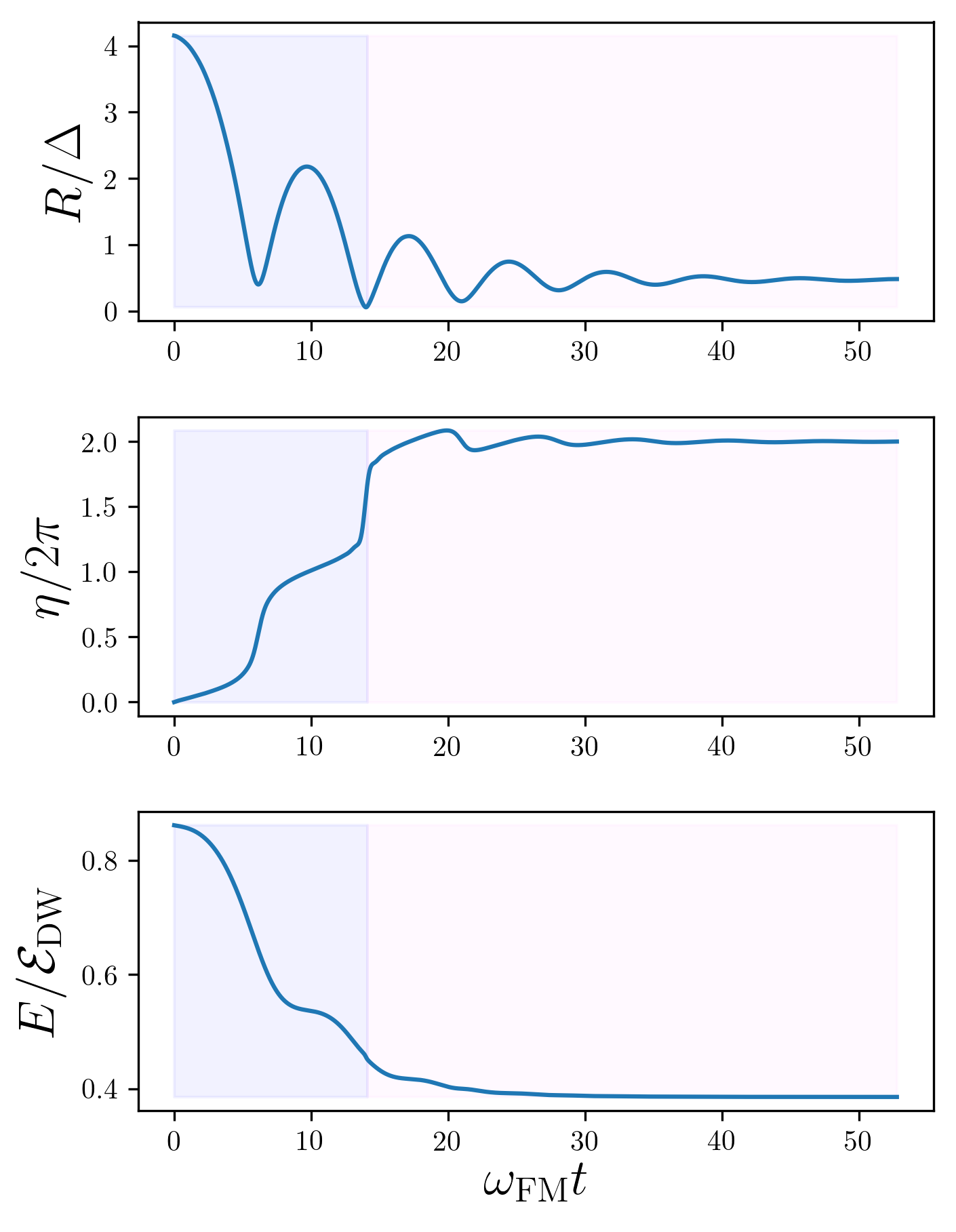}
    \caption{
    Transition from rotational breathing to oscillatory breathing from a micromagnetic simulation, showing one full rotation before traversing the saddle energy.
    The point where the energy drops below the energy saddle is estimated by closely inspecting the energy signal to find a sharp drop.
    The parameters in this simulation are $\alpha=0.1$, $D=2.8\times 10^{-3}$Jm$^{-2}$, $A=1.5\times 10^{-11}$ Jm$^{-1}$ and $K=1.1\times 10^{6}$ Jm$^{-3}$, i.e. $g=0.54$.
       }
    \label{fig:transition}
\end{figure}

All micromagnetic simulations reported in this paper were performed with an enhanced version of MicroMagnum.~\cite{MicroMagnum}

\subsection*{Simulation geometry details for Figs.\ 4--5}

All simulations used a mesh of uniformly-discretized cubic finite difference cells with periodic boundary conditions to approximate an infinite thin film system.
Figure~\ref{fig:high-energy-oscillation}(b) and (c) in the main text used $1024\times1024\times1$ nodes with cell length $0.25$nm.
Figure~\ref{fig:micromagnetic-matching} used,
\begin{itemize}
\item for (a): $2048\times2048\times1$ nodes with cell length $0.5$nm,
\item for (b): $512\times512\times1$ nodes with cell length $0.25$nm,
\item for (c): $256\times 256 \times 1$ nodes with cell length $0.25$m.
\end{itemize}

\subsection*{Calculation of $r$ and $\eta$}

For the calculation of the skyrmion radius in the simulations we used a linear interpolation between the finite difference cells where $m_{z}$ changes sign. 
For this we picked a fixed cut through the diameter of the skyrmion since it has no translational motion. 

For the calculation of $\eta$ we used a marching-squares algorithm to approximate the magnetization along the $m_{z}=0$ contour by using image processing tools,~\cite{VanderWalt2014} and then took a simple average of the $\eta$ values for the spins along these points, $\eta = (1/N)\sum_{i=1}^{N}\eta_{i}$, where $\eta_{i}=\arctan(m_{y,i}/m_{x,i})$ are the $N$ wall angles taken along the interpolated contour.

\subsection*{The transition from rotations to oscillations}

Observing a transition from the rotating-like breathing behavior and small oscillatory-like breathing in the micromagnetic simulations is non-trivial. 
This can be understood from the results reported in the main text as follows.
The amplitude of the breathing motion is larger for higher energies, so the total size of the magnetic material has to be sufficiently large.
At the same time  there also needs to be many finite difference cells because the trajectory of a breathing skyrmion typically passes close to the saddle point during the transition from the bowl region to the basin region and because the saddle point radius is always small, $r_{\mathrm{sad}}<1$.
In addition our model is intended to be more accurate for larger radius skyrmions, suggesting that best results should come for high coupling constants $|B|$. 
And yet we have the difficulty that $r_{\mathrm{sad}}\to 0^{+}$ as $|B|\to 1^{-}$. 
This presents a quandary: the more reliable we take the micromagnetic parameters in order to assess the model, the longer the simulations will last due to requiring more simulation cells to maintain numerical stability. 
Nonetheless, the transition can still be obtained, see Fig.~\ref{fig:transition}.

\end{document}